%% file: main.tex
\documentclass[a4paper,11pt]{article}
\usepackage{jcappub} % for details on the use of the package, please see the JINST-author-manual
\usepackage{lineno}
\usepackage{physics}
\usepackage{mydefs}
\usepackage[normalem]{ulem}
\usepackage{hyperref}
\usepackage{orcidlink}

\arxivnumber{2404.03117} % Only if you have one
\title{\boldmath Suppressing the sample variance of DESI-like galaxy clustering with fast simulations}

\input{authors}
\emailAdd{zhejied@sjtu.edu.cn}
\emailAdd{andrei.variu@epfl.ch}
\emailAdd{shadab.alam@tifr.res.in}

\abstract{Ongoing and upcoming galaxy redshift surveys, such as the Dark Energy Spectroscopic Instrument (DESI) survey, will observe vast regions of sky and a wide range of redshifts. In order to model the observations and address various systematic uncertainties, $N$-body simulations are routinely adopted, however, the number of large simulations with sufficiently high mass resolution is usually limited by available computing time. Therefore, achieving a simulation volume with the effective statistical errors significantly smaller than those of the observations becomes prohibitively expensive. In this study, we apply the Convergence Acceleration by Regression and Pooling (CARPool) method to mitigate the sample variance of the DESI-like galaxy clustering in the $\abacus$ simulations, with the assistance of the quasi-$N$-body simulations $\fastpm$. 
Based on the halo occupation distribution (HOD) models, we construct different $\fastpm$ galaxy catalogs, including the luminous red galaxies (LRGs), emission line galaxies (ELGs), and quasars, with their number densities and two-point clustering statistics well matched to those of $\abacus$. 
We also employ the same initial conditions between $\abacus$ and $\fastpm$ to achieve high cross-correlation, as it is useful in effectively suppressing the variance.
Our method of reducing noise in clustering is equivalent to performing a simulation with volume larger by a factor of $5$ and $4$ for LRGs and ELGs, respectively.
We also mitigate the standard deviation of the LRG bispectrum with the triangular configurations $k_2=2k_1=0.2\hMpc$ by a factor of 1.6.
With smaller sample variance on galaxy clustering, we are able to constrain the baryon acoustic oscillations (BAO) scale parameters to higher precision. The CARPool method will be beneficial to better constrain the theoretical systematics of BAO, redshift space distortions (RSD) and primordial non-Gaussianity (NG).}

\begin{document}
\maketitle
\flushbottom

\input{introduction}

\input{simulations}

\input{methodology}

\input{galaxy_clustering}

\input{result}

\input{conclusions}

\paragraph{Data availability.} We share all the necessary data and code to generate the figures and tables of this publication in zenodo repository https://zenodo.org/records/10644109.

\acknowledgments
We thank Johannes U. Lange, Joseph DeRose and the anonymous referee for their valuable comments and suggestions to improve the draft. We thank Eric Armengaud for the coordination of DESI internal review on the draft. ZD thanks Haojie Xu for helpful discussions on HOD. ZD and YY were supported by the National Key R\&D Program of China (2023YFA1607800, 2023YFA1607802), the National Science Foundation of China (grant numbers 12273020, 11621303, 11890691) and the science research grant from the China Manned Space Project with NO. CMS-CSST-2021-A03.
YY acknowledges the sponsorship from Yangyang Development Fund.
AV acknowledges support from the Swiss National Science Foundation (SNF) "Cosmology with 3D Maps of the Universe" research grant, 200020\_175751 and 200020\_207379.
SA acknowledges support of the Department of Atomic Energy, Government of india, under project no. 12-R\&D-TFR-5.02-0200. SA is partially supported by the European Research Council through the COSFORM Research Grant (\#670193) and STFC consolidated grant no. RA5496. 

This material is based upon work supported by the U.S. Department of Energy (DOE), Office of Science, Office of High-Energy Physics, under Contract No. DE–AC02–05CH11231, and by the National Energy Research Scientific Computing Center, a DOE Office of Science User Facility under the same contract. Additional support for DESI was provided by the U.S. National Science Foundation (NSF), Division of Astronomical Sciences under Contract No. AST-0950945 to the NSF’s National Optical-Infrared Astronomy Research Laboratory; the Science and Technology Facilities Council of the United Kingdom; the Gordon and Betty Moore Foundation; the Heising-Simons Foundation; the French Alternative Energies and Atomic Energy Commission (CEA); the National Council of Humanities, Science and Technology of Mexico (CONAHCYT); the Ministry of Science and Innovation of Spain (MICINN), and by the DESI Member Institutions: \url{https://www.desi.lbl.gov/collaborating-institutions}. Any opinions, findings, and conclusions or recommendations expressed in this material are those of the author(s) and do not necessarily reflect the views of the U. S. National Science Foundation, the U. S. Department of Energy, or any of the listed funding agencies.

The authors are honored to be permitted to conduct scientific research on Iolkam Du’ag (Kitt Peak), a mountain with particular significance to the Tohono O’odham Nation.

\input{appendix}

% Bibliography
%% [A] Recommended: using JHEP.bst file
\bibliographystyle{JHEP}
\bibliography{references.bib}

\end{document}

%% file: authors.tex
\author[1,2]{{Z.~Ding}\orcidlink{0000-0002-3369-3718},}
\author[3]{{A.~Variu}\orcidlink{0000-0001-8615-602X},}
\author[4]{{S.~Alam}\orcidlink{0000-0002-3757-6359},}
\author[1,2]{{Y.~Yu}\orcidlink{0000-0002-9359-7170},}
\author[5,6,7]{{C.~Chuang}\orcidlink{0000-0002-3882-078X},}
\author[8,9]{{E.~Paillas}\orcidlink{0000-0002-4637-2868},}
\author[10]{{C.~Garcia-Quintero}\orcidlink{0000-0003-1481-4294},}
\author[11]{{X.~Chen},}
\author[12]{{J.~Mena-Fern\'andez}\orcidlink{0000-0001-9497-7266},}
\author[13]{{J.~Aguilar},}
\author[14]{{S.~Ahlen}\orcidlink{0000-0001-6098-7247},}
\author[15]{{D.~Brooks},}
\author[13]{{T.~Claybaugh},}
\author[16]{{A.~de la Macorra}\orcidlink{0000-0002-1769-1640},}
\author[15]{{P.~Doel},}
\author[17,6]{{K.~Fanning}\orcidlink{0000-0003-2371-3356},}
\author[18,19]{{J.~E.~Forero-Romero}\orcidlink{0000-0002-2890-3725},}
\author[20,21,22]{{E.~Gaztañaga},}
\author[13]{{S.~Gontcho A Gontcho}\orcidlink{0000-0003-3142-233X},}
\author[23]{{G.~Gutierrez},}
\author[24]{{C.~Hahn}\orcidlink{0000-0003-1197-0902},}
\author[25,26,27]{{K.~Honscheid},}
\author[28]{{C.~Howlett}\orcidlink{0000-0002-1081-9410},}
\author[29]{{S.~Juneau},}
\author[30]{{R.~Kehoe},}
\author[13]{{T.~Kisner}\orcidlink{0000-0003-3510-7134},}
\author[13]{{A.~Kremin}\orcidlink{0000-0001-6356-7424},}
\author[13]{{A.~Lambert},}
\author[13]{{M.~Landriau}\orcidlink{0000-0003-1838-8528},}
\author[31]{{L.~Le~Guillou}\orcidlink{0000-0001-7178-8868},}
\author[32,33]{{M.~Manera}\orcidlink{0000-0003-4962-8934},}
\author[34,33]{{R.~Miquel},}
\author[35]{{E.~Mueller},}
\author[36]{{A.~D.~Myers},}
\author[37]{{J.~Nie}\orcidlink{0000-0001-6590-8122},}
\author[38,39]{{G.~Niz}\orcidlink{0000-0002-1544-8946},}
\author[13,40,41]{{C.~Poppett},}
\author[42]{{M.~Rezaie}\orcidlink{0000-0001-5589-7116},}
\author[43]{{G.~Rossi},}
\author[44]{{E.~Sanchez}\orcidlink{0000-0002-9646-8198},}
\author[45,46]{{M.~Schubnell},}
\author[47]{{H.~Seo}\orcidlink{0000-0002-6588-3508},}
\author[13]{{J.~Silber}\orcidlink{0000-0002-3461-0320},}
\author[29]{{D.~Sprayberry},}
\author[46]{{G.~Tarl\'{e}}\orcidlink{0000-0003-1704-0781},}
\author[16]{{M.~Vargas-Maga\~na}\orcidlink{0000-0003-3841-1836},}
\author[37]{{H.~Zou}\orcidlink{0000-0002-6684-3997},}

\affiliation[1]{Department of Astronomy, School of Physics and Astronomy, Shanghai Jiao Tong University, Shanghai 200240, China}
\affiliation[2]{Key Laboratory for Particle Astrophysics and Cosmology (MOE)/Shanghai Key Laboratory for Particle Physics and Cosmology, Shanghai 200240, China}
\affiliation[3]{Ecole Polytechnique F\'{e}d\'{e}rale de Lausanne, CH-1015 Lausanne, Switzerland}
\affiliation[4]{Tata Institute of Fundamental Research, Homi Bhabha Road, Mumbai 400005, India}
\affiliation[5]{Department of Physics and Astronomy, The University of Utah, 115 South 1400 East, Salt Lake City, UT 84112, USA}
\affiliation[6]{Physics Department, Stanford University, Stanford, CA 93405, USA}
\affiliation[7]{SLAC National Accelerator Laboratory, Menlo Park, CA 94305, USA}
\affiliation[8]{Department of Physics and Astronomy, University of Waterloo, 200 University Ave W, Waterloo, ON N2L 3G1, Canada}
\affiliation[9]{Waterloo Centre for Astrophysics, University of Waterloo, 200 University Ave W, Waterloo, ON N2L 3G1, Canada}
\affiliation[10]{Department of Physics, The University of Texas at Dallas, Richardson, TX 75080, USA}
\affiliation[11]{Physics Department, Yale University, P.O. Box 208120, New Haven, CT 06511, USA}
\affiliation[12]{Laboratoire de Physique Subatomique et de Cosmologie, 53 Avenue des Martyrs, 38000 Grenoble, France}
\affiliation[13]{Lawrence Berkeley National Laboratory, 1 Cyclotron Road, Berkeley, CA 94720, USA}
\affiliation[14]{Physics Dept., Boston University, 590 Commonwealth Avenue, Boston, MA 02215, USA}
\affiliation[15]{Department of Physics \& Astronomy, University College London, Gower Street, London, WC1E 6BT, UK}
\affiliation[16]{Instituto de F\'{\i}sica, Universidad Nacional Aut\'{o}noma de M\'{e}xico,  Cd. de M\'{e}xico  C.P. 04510,  M\'{e}xico}
\affiliation[17]{Kavli Institute for Particle Astrophysics and Cosmology, Stanford University, Menlo Park, CA 94305, USA}
\affiliation[18]{Departamento de F\'isica, Universidad de los Andes, Cra. 1 No. 18A-10, Edificio Ip, CP 111711, Bogot\'a, Colombia}
\affiliation[19]{Observatorio Astron\'omico, Universidad de los Andes, Cra. 1 No. 18A-10, Edificio H, CP 111711 Bogot\'a, Colombia}
\affiliation[20]{Institut d'Estudis Espacials de Catalunya (IEEC), 08034 Barcelona, Spain}
\affiliation[21]{Institute of Cosmology and Gravitation, University of Portsmouth, Dennis Sciama Building, Portsmouth, PO1 3FX, UK}
\affiliation[22]{Institute of Space Sciences, ICE-CSIC, Campus UAB, Carrer de Can Magrans s/n, 08913 Bellaterra, Barcelona, Spain}
\affiliation[23]{Fermi National Accelerator Laboratory, PO Box 500, Batavia, IL 60510, USA}
\affiliation[24]{Department of Astrophysical Sciences, Princeton University, Princeton NJ 08544, USA}
\affiliation[25]{Center for Cosmology and AstroParticle Physics, The Ohio State University, 191 West Woodruff Avenue, Columbus, OH 43210, USA}
\affiliation[26]{Department of Physics, The Ohio State University, 191 West Woodruff Avenue, Columbus, OH 43210, USA}
\affiliation[27]{The Ohio State University, Columbus, 43210 OH, USA}
\affiliation[28]{School of Mathematics and Physics, University of Queensland, 4072, Australia}
\affiliation[29]{NSF NOIRLab, 950 N. Cherry Ave., Tucson, AZ 85719, USA}
\affiliation[30]{Department of Physics, Southern Methodist University, 3215 Daniel Avenue, Dallas, TX 75275, USA}
\affiliation[31]{Sorbonne Universit\'{e}, CNRS/IN2P3, Laboratoire de Physique Nucl\'{e}aire et de Hautes Energies (LPNHE), FR-75005 Paris, France}
\affiliation[32]{Departament de F\'{i}sica, Serra H\'{u}nter, Universitat Aut\`{o}noma de Barcelona, 08193 Bellaterra (Barcelona), Spain}
\affiliation[33]{Institut de F\'{i}sica d’Altes Energies (IFAE), The Barcelona Institute of Science and Technology, Campus UAB, 08193 Bellaterra Barcelona, Spain}
\affiliation[34]{Instituci\'{o} Catalana de Recerca i Estudis Avan\c{c}ats, Passeig de Llu\'{\i}s Companys, 23, 08010 Barcelona, Spain}
\affiliation[35]{Department of Physics and Astronomy, University of Sussex, Brighton BN1 9QH, U.K}
\affiliation[36]{Department of Physics \& Astronomy, University  of Wyoming, 1000 E. University, Dept.~3905, Laramie, WY 82071, USA}
\affiliation[37]{National Astronomical Observatories, Chinese Academy of Sciences, A20 Datun Rd., Chaoyang District, Beijing, 100012, P.R. China}
\affiliation[38]{Departamento de F\'{i}sica, Universidad de Guanajuato - DCI, C.P. 37150, Leon, Guanajuato, M\'{e}xico}
\affiliation[39]{Instituto Avanzado de Cosmolog\'{\i}a A.~C., San Marcos 11 - Atenas 202. Magdalena Contreras, 10720. Ciudad de M\'{e}xico, M\'{e}xico}
\affiliation[40]{Space Sciences Laboratory, University of California, Berkeley, 7 Gauss Way, Berkeley, CA  94720, USA}
\affiliation[41]{University of California, Berkeley, 110 Sproul Hall \#5800 Berkeley, CA 94720, USA}
\affiliation[42]{Department of Physics, Kansas State University, 116 Cardwell Hall, Manhattan, KS 66506, USA}
\affiliation[43]{Department of Physics and Astronomy, Sejong University, Seoul, 143-747, Korea}
\affiliation[44]{CIEMAT, Avenida Complutense 40, E-28040 Madrid, Spain}
\affiliation[45]{Department of Physics, University of Michigan, Ann Arbor, MI 48109, USA}
\affiliation[46]{University of Michigan, Ann Arbor, MI 48109, USA}
\affiliation[47]{Department of Physics \& Astronomy, Ohio University, Athens, OH 45701, USA}

%% file: introduction.tex
\section{Introduction}
\label{sec:intro}
The Dark Energy Spectroscopic Instrument (DESI) is an ongoing Stage IV spectroscopic redshift survey \cite{DESI2016a, DESI2016b, DESI_instrument_2022, Adame2023a}. DESI will cover a large sky area $\sim 14000$ deg$^2$ and a wide redshift range $0<z<3.5$, targeting different tracers of dark matter distribution, i.e. the bright galaxy samples (BGS) at $0<z<0.4$ \citep{Hahn2023}, luminous red galaxies (LRGs) at $0.4<z<1.1$ \citep{Zhou2023}, emission line galaxies (ELGs) at $1.1<z<1.6$ \citep{Raichoor2023}, quasars (QSOs) at $0.9<z<2.1$ \citep{Chaussidon2023}, and Lyman alpha (Ly-$\alpha$) forest at $2.1<z<3.5$. At the end of survey, DESI is going to collect over $40$ million extra-galactic spectra of galaxies and quasars, one order of magnitude larger than the spectra from the previous galaxy redshift surveys, e.g. the 2-degree Field Galaxy Redshift Survey \citep{Cole2005}, the WiggleZ Dark Energy Survey \citep{Blake2011b}, the 6-degree Field Galaxy Survey \citep{Beutler2011}, and the Sloan Digital Sky Survey \citep{Eisenstein2005, Anderson2014, Beutler2017, Bautista2017, Bourboux2020, Wang2020, Zhao2021} combined together. Taking the baryon acoustic oscillations (BAO) as a standard ruler, DESI is able to measure the cosmological distances at sub-per cent level \citep{Adame2023a}, which dramatically tightens the constraints on the expansion rate of the Universe and the dark energy models. Meanwhile, DESI measures the redshift space distortions (RSD), which is originated from the peculiar motions of galaxies. RSD adds anisotropy on the measured galaxy clustering signal \citep{Kaiser1987, Hamilton1998}. Measuring RSD can directly probe the structure growth rate and the amount of matter in the Universe, hence, it is essential to constrain gravity models \citep{Beutler2014, Hector2020, Zhao2021}. In addition, we can probe primordial non-Gaussianity (NG) from high-order galaxy clustering statistics, such as bispectrum \citep{Verde2001, Scoccimarro2004}. For the case of local NG, it would induce a scale-dependent galaxy bias proportional to $k^{-2}$ on galaxy power spectrum \citep{Dalal2008}. Several studies have been conducted recently to constrain the local NG parameter $\fnl$, e.g. \cite{Castorina2019, Mueller2022, Cabass2022, D'Amico2022, Rezaie2023}.

Besides DESI, other forthcoming Stage IV galaxy surveys including the \textit{Euclid} space telescope \cite{Euclid2011}, the Prime Focus Spectrograph \cite{Takada2014}, the China Space Station Telescope \cite{Gong2019}, and the Nancy Grace Roman Space Telescope \cite{Spergel2015}, will dramatically increase the survey volumes and sample size. 
In order to interpret the observation and calibrate different systematic errors for cosmological analysis, we widely refer to $N$-body simulations. In DESI, the flagship $N$-body simulation is $\abacus$ \citep{Maksimova2021}.\footnote{https://abacussummit.readthedocs.io/en/latest/simulations.html} Due to its large volume and high mass resolution, the number of $\abacus$ simulations is limited, e.g. 25 base boxes for the primary cosmology. 
But for other cosmologies the number of simulations is less, and there is only one simulation for most of cosmologies. Comparing the simulation box size $8\Gpchcube$ with the DESI effective survey volume $\sim 20\Gpchcube$ \citep{Adame2023a}, the sample variance of simulations will be larger than that of DESI data.

There have been several methods proposed to mitigate the sample variance of simulations. \cite{Angulo2016} proposed the fixed and paired method, which utilizes a paired initial conditions (ICs) with the fixed amplitude and inverse phases. With a small number of the paired simulations, it can significantly suppress the sample variance of dark matter, halo and galaxy two-point clustering statistics on large scales without introducing bias \citep[e.g.][]{Villaescusa-Navarro2018, Chuang2019, Maion2022}. It has been applied and studied in various simulations, e.g. \citep{Avila2023, Acharya2023, Hernandez2023}.
Another method recently proposed is adopting the principle of control variates, including the simulation based one and the theory based one. The former utilizes fast simulations or surrogates to construct the control variates \citep{Chartier2021, Chartier2022, Ding2022, Lee2024}, while the latter theoretically predicts the control variates \citep{Kokron2022, DeRose2023a, DeRose2023b, Hadzhiyska2023b}. The analytic control variates only require ICs without running a number of surrogates as needed for the simulation based one, hence, it can save quite amount of computational time. However, currently it is only available to model galaxy two-point clustering but not for higher order statistics, e.g. bispectrum. In addition, it is not trivial to include some observational systematics, such as fiber collision, in the theoretical template, while it is straightforward for simulations. In this study, we focus on the simulation based method, and study the performance of the sample variance suppression on galaxy clustering.  

Galaxies form in dark matter halos and are tracers of dark matter distribution. In cosmological simulations with dark matter only, we directly simulate the spatial distribution and motion of dark matter particles. Then we define some gravitational bounded regions with particles concentrated as dark matter halos. After that, we generate the galaxy distribution by painting galaxies into halos based on some galaxy-halo connection schemes varying from more physical to more empirical ones (see a recent review \citep{Wechsler2018}). The halo occupation distribution (HOD) is a phenomenological model \citep{Jing1998, Peacock2000, Zheng2005}, which models the distribution of central and satellite galaxies separately. 
In this study, we apply the HOD models to generate $\fastpm$ galaxy catalogs, whose number density and galaxy clustering are matched to those of $\abacus$. 

Our paper is constructed as follows. In section \ref{sec:simulation}, we summarize the details of $\abacus$ and $\fastpm$ simulations. In section \ref{sec:method}, we introduce the HOD models, the HOD fitting process, and the CARPool method. In section \ref{sec:clustering}, we describe the galaxy clustering statistics that we study. In section \ref{sec:result}, we show the comparison of the $\fastpm$ and $\abacus$ galaxy clustering, the suppression on the sample variance of the $\abacus$ galaxy clustering and the increased BAO constraints from CARPool. In section \ref{sec:conclusion}, we make conclusions and discussions.

%% file: simulations.tex
\section{Simulations}\label{sec:simulation}
\subsection{AbacusSummit}
\textsc{AbacusSummit} is a large suite of high-accuracy $N$-body simulations prepared for the DESI survey. The simulations were generated at the Summit supercomputer of the Oak Ridge National Laboratory using the \textsc{Abacus} code \cite{Garrison2021}. $\abacus$ covers different cosmologies. In our study, we focus on the simulations with the primary cosmology, denoted as \texttt{c000}, which is the $\Lambda$CDM model with the cosmological parameters constrained from the Planck 2018 result (TT,TE,EE+lowE+lensing) \cite{Planck2018}. $\abacus$ also covers different box sizes and mass resolutions of dark matter particles. There are 25 \texttt{base} boxes with the cosmology \texttt{c000} but different initial conditions (ICs), denoted as \texttt{ph0[00-24]}. Each realization has the box size $2\Gpch$ per side and contains $6912^3$ particles with mass equal to $\sim 2.1\times 10^9\Msunh$. 
$\abacus$ utilizes \textsc{compaso} to identify dark matter halos and adopts a cleaning method to remove spuriously identified halos \cite{Hadzhiyska2022}. Based on the cleaned halo catalogs at different redshifts, different types of galaxy catalogs have been generated using HOD. We describe the HOD models in section \ref{sec:method_hod}. The best-fit HOD parameters are obtained from fitting the $\abacus$ galaxy clustering to the observed one from the DESI One-Percent Survey\footnote{The HOD fitting pipeline is based on \textsc{abacushod}, which is a part of \textsc{abacusutils} https://abacusutils.readthedocs.io/en/latest/ \cite{Yuan2022}. The One-Percent Survey is the final phase of the DESI survey validation (SV), known as SV3. We utilize the first generation of $\abacus$ galaxy catalogs, whose HOD parameters are based on the early version of SV3.} \citep{Adame2023b}. Therefore, the obtained galaxy catalogs with the best-fit HOD have similar galaxy clustering as the true one that DESI will observe, ignoring any effects from survey footprint, observational systematics, and fiber assignment. We call these galaxy catalogs as DESI-like samples. Table \ref{tab:cubic_mock_info} displays the basic information of the DESI-like catalogs for different tracers.

\begin{table}
    \centering
    \begin{tabular}{|c | c | c | c| c |}
        \hline
          & redshift & $n$ [$10^{-4}\hMpccube$] & bias & $f$ \\
         \hline
         %BGS & 0.2 & 5 & 1.6 & 0.636 \\
         LRGs & 0.8 & 10 & 2.0 & 0.838 \\
         ELGs & 1.1 & 30 & 1.2 & 0.888\\
         QSOs & 1.4 & 1.3 & 2.1 & 0.920\\
         \hline
    \end{tabular}
    \caption{The redshift, galaxy number density, galaxy bias and growth rate of the cubic DESI-like galaxy catalogs from \textsc{AbacusSummit}.}
    \label{tab:cubic_mock_info}
\end{table}

\subsection{\textsc{FastPM}}
\textsc{FastPM}\footnote{https://github.com/fastpm/fastpm} is a type of quasi-$N$-body simulations \citep{Feng2016}. It implements the Particle-Mesh (PM) scheme with modified kick and drift factors to ensure that the linear growth of displacement agrees with the Zel'dovich approximation at large scales. With a lower mass resolution compared to the $N$-body simulation (e.g. $\abacus$) and a few time steps ($\sim 40$), $\fastpm$ is able to recover the matter power spectrum up-to $k\sim 1\hMpc$ within $2$ per cent level accuracy \citep{Grove2022}. $\fastpm$ uses Friends-of-Friends (FoF) algorithm to find dark matter halos. In addition, several methods have been proposed to further improve the $\fastpm$ matter and halo clustering at small scales \cite{Dai2018, Dai2020}. Meanwhile, \cite{Bayer2021} implements the contribution of neutrinos in $\fastpm$. \cite{Variu2023} populates galaxies in $\fastpm$ halos based on HOD, and makes the galaxy clustering well matched to that of $\abacus$.

In our study, we utilize the $\fastpm$ simulations generated in \cite{Ding2022}, which contains $25$ realizations with the cosmology \texttt{c000} and the same ICs as the $\abacus$ \texttt{base} boxes. In addition, we have generated $313$ boxes with random (nonfixed-amplitude) ICs. The simulation box has size of $2\Gpch$ per side and $5184^3$ particles with mass resolution $\sim 5\times 10^9\Msunh$. The force resolution is set by the parameter $B=N_{m}/N_g=2$, where $N_m$ and $N_g$ are the number of the mesh size and galaxies per side of box, respectively. Simulations start from the initial redshift $19$ to $0.1$ with $40$ time steps linearly separated in scale $a=\frac{1}{1+z}$. One simulation takes about $50$ minutes using 1152 KNL nodes each with 68 CPU cores and 98 GB memory based on the Cori\footnote{The Cori supercomputer has retired in May 2023.} supercomputer at NERSC\footnote{https://www.nersc.gov}. Although the computational cost is much cheaper than that of $N$-body simulation, it is still costly to run a number of $\fastpm$ with such configuration. To populate galaxies in $\fastpm$ halos, we select halos with mass larger than $5\times 10^{10}\Msunh$. The particle mass resolution in our simulation is close to the low resolution one of \citep{Variu2023}, which uses a halo mass cut similar to ours, and gives reliable HOD fitting. We have also tested using a higher halo mass cut, and found that it does not affect our final result.

%% file: methodology.tex
\section{Methodology}\label{sec:method}
\subsection{HOD models}\label{sec:method_hod}
We describe the HOD models used to populate different types of galaxies in the $\abacus$ and $\fastpm$ catalogs. We also notify some modifications on the $\fastpm$ HOD models.

\subsubsection{LRGs}
At $0.4<z<1.1$, DESI mainly observes LRGs, which are relatively easy to select with their characteristic 4000 $\mathring{A}$ break in the spectra. They are assumed to inhabit massive halos and have high galaxy biases. We adopt the vanilla HOD model based on \cite{Zheng2005} for $\abacus$ LRGs. In this model, the distribution of central galaxies follows a Bernoulli distribution with the mean occupation number given by
\begin{equation}
    \langle N_{\text{cent}}^{\text{LRG}}\rangle = \frac{1}{2} \bigg[1+\text{erf}\Big(\frac{\log M_{\text{h}} - \log M_{\text{cut}}}{\sqrt{2} \sigma_{\log M_{\text{h}}}}\Big)\bigg],
\end{equation}
where $M_{\text{h}}$ is the halo mass, $M_{\text{cut}}$ is the halo mass limit at which half of the halos host one central galaxy on average, $\log$ denotes the 10 based logarithm, and erf is the error function, i.e.
\begin{align}
\text{erf}(x)=\frac{2}{\sqrt{\pi}}\int_0^x e^{-t^2} dt.
\end{align}
The distribution of satellite galaxies follows a Poisson distribution. The mean number distribution follows a power low, i.e. 
\begin{align}
    \langle N_{\text{sat}}^{\text{LRG}}\rangle = \bigg(\frac{M_{\text{h}}-\kappa M_{\text{cut}}}{M_1}\bigg)^{\alpha},\label{eq:sate_hod}
\end{align}
where $\kappa M_{\text{cut}}$ denotes the minimum mass that halos can host satellite galaxies, $M_1$ is the halo mass at which halos host on average one satellite galaxy, and $\alpha$ is the slope of the power-law function. For the spatial distribution of satellites, it follows the Navarro-Frenk-White profile \cite{NFW1996}. We assign velocities to satellites based on a Gaussian distribution with the mean halo velocity and the velocity distribution of dark matter particles inside a halo.

To make the $\fastpm$ galaxy clustering match with that of $\abacus$ better in redshift space, we add an additional parameter $\vdisp$ on top of the default model \citep{Variu2023}. It modifies velocities along the line of sight (LoS)\footnote{For a cubic box, we fix the line of sight along the $z$ axis.} for the $\fastpm$ satellites, i.e.
\begin{align}
    v^{\text{sat, modified}}_{\parallel} = (v^{\text{sat, default}}_{\parallel} - v^{\text{halo}}_{\parallel}) \times \vdisp + v^{\text{halo}}_{\parallel},
\end{align}
where the subscript $\parallel$ denotes parallel to the LoS. In this study, we include $\vdisp$ in all the $\fastpm$ HOD models.

\subsubsection{ELGs}
At $0.8<z<1.6$, DESI observes the primary galaxy samples with strong [O$\,$II] doublet $\lambda\lambda$ 3726,29 $\mathring{A}$ emission lines, denoted as ELGs, which are sub-samples of the star-forming galaxies. Star forming process depends on halo mass and environment. For central ELGs living in more massive halos, they are more likely to quench, hence, the canonical step function cannot fully describe the distribution.
The high mass quenching model \cite{Gonzalez-Perez2018, Avila2020, Alam2020} works quite well for central ELGs, and it models the mean number distribution as
\begin{align}
    \langle N_{\text{cen}}^{\text{ELG}}\rangle = 2A \phi(M)\Phi(\gamma M) + \frac{1}{2Q}\bigg[1+\text{erf}\bigg(\frac{\text{log}M_h - \text{log}M_{\text{cut}}}{0.01}\bigg)\bigg]. \label{eq:elg_hod}
\end{align}
where
\begin{align}
\phi(x) &= \mathcal{N}(\log M_{\text{cut}}, \sigma_{\text{log} M_{\text{h}}}),\\
\Phi(x) &= \int_{-\infty}^{x} \phi(x')dx' = \frac{1}{2}\bigg[1+\text{erf}\bigg(\frac{x}{\sqrt{2}}\bigg)\bigg],\\
A &= \frac{p_{\max} - 1/Q}{\text{max}(2\phi(x)\Phi(\gamma x))}. 
\end{align}
% \begin{align}
%     \langle N_{\text{cen}}^{\text{ELG}}\rangle = A \, \mathcal{N}(\log M_{\text{h}}, \sigma_{\log M_{\text{h}}}) \bigg(1 + \text{erf} \Big(\gamma \frac{\log M_{\text{h}} - \log M_{\text{cut}}}{\sqrt{2} \sigma_{\log M_{\text{h}}}} \Big)\bigg),
% \end{align}
$\mathcal{N}$ denotes the Gaussian distribution, $Q$ models the galaxy quenching efficiency at massive halos, $p_{\text{max}}$ sets the saturation level of the occupation probability, and $\gamma$ is a skewness parameter. We refer the readers to \cite{Alam2020} for more detailed study on these parameters. For the $\fastpm$ centrals, we only adopt the first part of eq. \ref{eq:elg_hod} to populate. For the $\abacus$ ELG satellites, the mean number distribution follows the same relation as the LRGs' (eq. \ref{eq:sate_hod}). Recent studies have found that the galactic conformity has strong effect on the population of satellites. Satellites are more likely to inhabit halos with central galaxies as ELGs \cite{Hadzhiyska2023a, Gao2023, Rocher2023}. Such galactic conformity can boost the one-halo term clustering at very small scales. We leave the study of galactic conformity in future work.

\subsubsection{QSOs}
QSOs are luminous point-like sources, generated by the accretion of matter onto super-massive black holes in centers of galaxies. DESI observes QSOs as direct tracers at $0.9<z<2.1$, and takes them as backlights for the Ly $\alpha$ forest at $2.1<z<3.6$. Here, we only consider the QSOs as direct tracers. For the central distribution of $\abacus$ QSOs, it is the same as the LRGs' but with the downsampling parameter $p_{\text{max}}$ that models the duty cycle of QSOs \citep{Alam2020, Yuan2023}, i.e.  
\begin{align}
    \langle N_{\text{cent}}^{\text{QSO}}\rangle = \frac{1}{2} p_{\text{max}}\bigg[1+\text{erf}\Big(\frac{\log M_{\text{h}} - \log M_{\text{cut}}}{\sqrt{2} \sigma_{\log M_{\text{h}}}}\Big)\bigg],
\end{align}
For the distribution of satellites, it follows eq. \ref{eq:sate_hod} as well.

\subsection{HOD fitting process}
\begin{figure}
    \centering
    \includegraphics[width=0.98\linewidth]{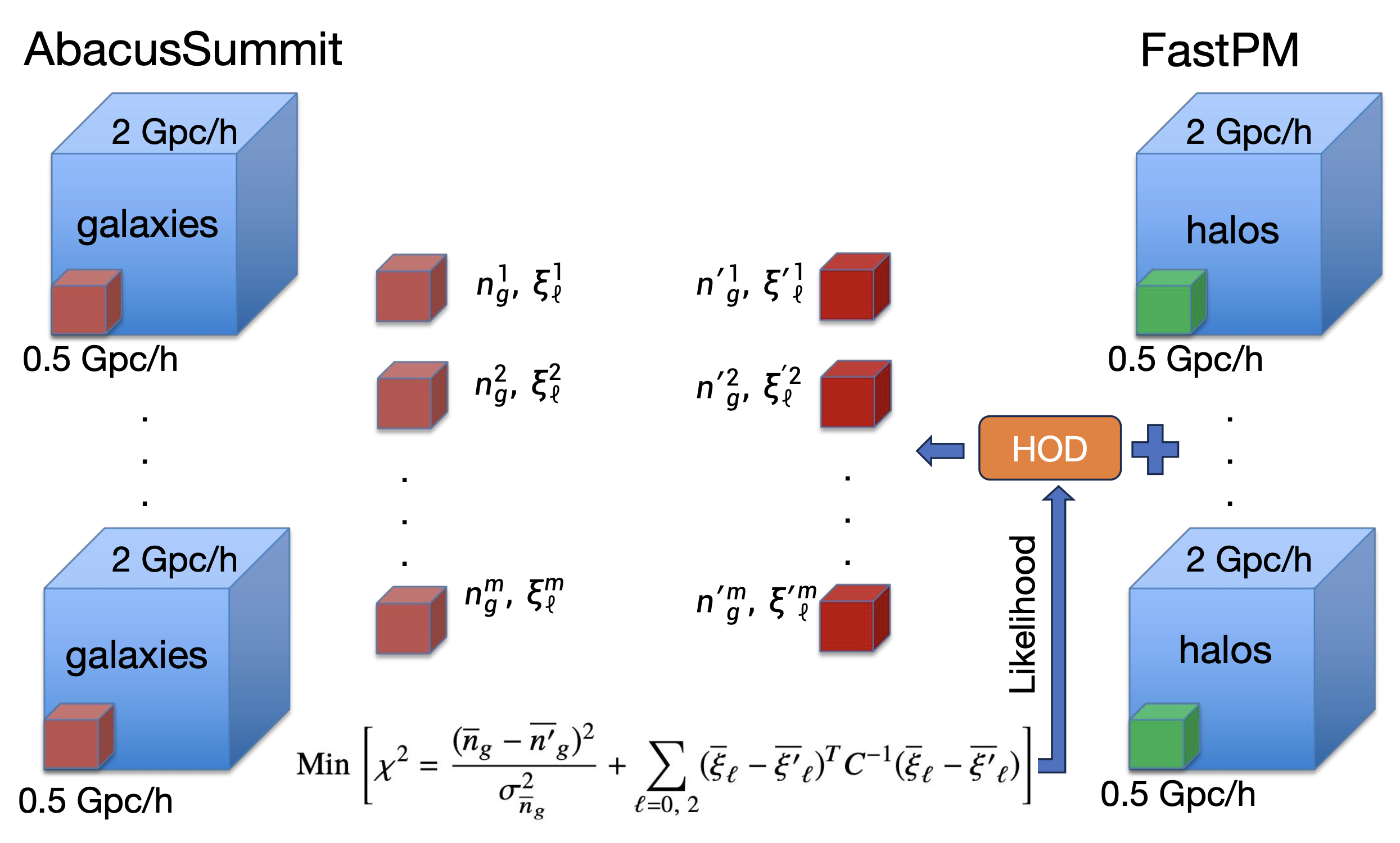}    \caption{A cartoon diagram showing the flowchart of the HOD fitting process. Given an $\abacus$ galaxy catalog with the original box size $(2\Gpch)^3$, we cut out a $500\Mpch$ sub-box. For the sub-box, we calculate the galaxy number density and two-point clustering statistics. From a paired $\fastpm$ simulation with the $\abacus$ IC, we cut out the same sub-box, and populate galaxies in the halos given a HOD model. To find the optimal HOD parameters, we compare the galaxy number density and two-point clustering, and minimize the difference between the two galaxy catalogs. In order to reduce the sample variance, we compare the mean galaxy number density and clustering over a number of the paired sub-boxes. We use the nested sampling to find the best-fit HOD parameters.}
    \label{fig:hod_fit}
\end{figure}
To find the optimal HOD parameters, we follow the routine described in \cite{Variu2023} with some modifications. Basically, we match the $\fastpm$ galaxy number density $\ngal$ and galaxy two-point clustering $\xi_\ell$ (or $P_\ell$) to those of $\abacus$. We find the best $\fastpm$ HOD parameters via minimizing
\begin{align}\label{eq:chi2_fit}
    \chi^2 = (D - M)^T \Sigma^{-1} (D - M),
\end{align}
where $D$ and $M$ denotes data (observation) and model prediction, respectively, and $\Sigma$ is the covariance matrix of $D$. 

Since we use the paired $\fastpm$ with the $\abacus$ ICs, some of the sample variance can be cancelled out when we compare the galaxy statistics from the two simulations, especially at larger scales. We consider the data as the difference of the galaxy number density and clustering from $\fastpm$ and $\abacus$, i.e. $D\rightarrow D_\text{F} - D_\text{A}$ with the subscripts F and A denoting $\fastpm$ and $\abacus$, respectively.
Ideally, we expect that $D_\text{F}$ can match closely to $D_\text{A}$, given some optimal HOD parameters, hence, the model expectation of $D_\text{F} - D_\text{A}$ is close to 0, i.e. $M\rightarrow 0$.    
We can transform eq.~\ref{eq:chi2_fit} to
\begin{align}
    \chi^2 \simeq (D_\text{F} - D_\text{A})^T \Sigma_{\text{diff}}^{-1} (D_\text{F} - D_\text{A}),
\end{align}
where $\Sigmadiff$ is the covariance matrix of $D_\text{F} - D_\text{A}$. However, we do not know $\Sigmadiff$ without having \textsc{FastPM} galaxy catalogs. 
To resolve that, our fitting process consists of two steps. 

For the first-step fitting, we aim to get initial guess on the $\fastpm$ HOD, so that it more or less reproduces the $\abacus$ clustering. This will allow us to obtain a first estimate of the difference covariance matrix $\Sigmadiff$. We then use such difference covariance matrix to run the second step of fitting and obtain the final HOD parameter for the $\fastpm$ simulations. 

In the first step, we consider the $\abacus$ galaxy statistics as the model, to which we fit the $\fastpm$ statistics. Instead of using the original $2\Gpch$ boxes, we choose sub-boxes to perform HOD fitting for two reasons. Firstly, it is computationally more efficient to populate galaxies and calculate the clustering given a set of sub-boxes thanks to the smaller volume. It is especially necessary for the case when we need to perform such process multiple times during the HOD fitting with some sampling method. Secondly, we can construct a covariance matrix $\Sigma$ from the sub-boxes. In our case, we cut each $\abacus$ box into 64 sub-boxes, each of which has side length $500\Mpch$. We compute the galaxy number density and clustering for each sub-box, and get the covariance matrix from 1600 ($25\times64$) sub-boxes. Given that in this step we just aim to estimate the $\fastpm$ HOD parameters, we simply use the diagonal terms of the covariance matrix, i.e. 
\begin{align}
    \Sigma= \text{Diag}\bigg[\sigma^2_{\xi_{1}(s_1)}\; \sigma^2_{\xi_{1}(s_2)} \; \cdots \; \sigma^2_{\xi_{1}(s_N)} \; \sigma^2_{\xi_{2}(s_1)}\, \sigma^2_{\xi_{2}(s_2)} \; \cdots \; \sigma^2_{\xi_{2}(s_N)}\; \sigma^2_{\nbargal} \bigg],\label{eq:Sigma}
\end{align}
where we consider the correlation function monopole and quadrupole, as well as the galaxy number density. We set $s_1=2\Mpch$ and $s_N=46\Mpch$ with the step size $4\Mpch$ for the coordinates of the correlation function multipoles. $\sigma^2_{\nbargal}$ denotes the sample variance of the mean galaxy number density, i.e. $\sigma^2_{\nbargal}=\sigma^2_{\ngal}/N_{\text{fit}}$ with $N_{\text{fit}}$ as the number of mocks that we fit. In our case, we use $N_{\text{fit}}=16$ sub-boxes, which are from different $2\Gpch$ boxes. As suggested in \citep{Variu2023}, the scale factor $1/N_{\text{fit}}$ up-weights the match of the galaxy number density with the reference in the HOD fitting. It can boost the fitting convergence, since we only calculate the $\fastpm$ galaxy clustering and compare it with the $\abacus$ one if the mean $\fastpm$ galaxy number density is less than $10\sigma_{\nbargal}$ different from the reference. It works quite well for the LRG and ELG HOD fitting. However, for QSOs with a much lower galaxy number density, the clustering signal becomes noisy with larger error bars. We find that it is better to add $1/N_{\text{fit}}$ as well for the sample variance of clustering statistics in eq. \ref{eq:Sigma} to improve the HOD fitting.
 
For $\fastpm$, we cut out the sub-boxes with the same way as $\abacus$. Given some HOD model, we can populate the $\fastpm$ halos with galaxies in the sub-boxes. To reduce the sample variance, we calculate the mean galaxy number density and two-point clustering from $N_{\text{fit}}$ sub-boxes, and compare them with those from the paired $\abacus$ sub-boxes. 
Minimizing the difference between the two based on eq. \ref{eq:chi2_fit}, we can find the best-fit HOD parameters. We use the fitting pipeline \textsc{hodor},\footnote{https://github.com/Andrei-EPFL/HODOR.} which implements \textsc{halotools}\footnote{https://halotools.readthedocs.io/en/latest/index.html} \citep{Hearin2017}, and utilizes \textsc{PyMultiNest} \citep{Feroz2008, Feroz2009, Feroz2019, Buchner2014} as a nested sampling Monte Carlo library. We show the demonstration of the HOD fitting process in figure \ref{fig:hod_fit}.

For the second-step fitting, we first obtain the $\fastpm$ galaxy catalogs based on the best-fit HOD parameters from the first-step fitting. Then we cut 25 $\fastpm$ ($2\Gpch$) boxes into 1600 ($500\Mpch$) sub-boxes. We calculate the covariance matrix of the difference of the two-point galaxy clustering ($D_\text{A}-D_\text{F}$) between $\abacus$ and $\fastpm$, denoted as $\mathbb{C}$. We concatenate $\mathbb{C}$ with the sample variance of the $\abacus$ galaxy number density to get

\begin{align}
    \Sigma_{\text{diff}}=
\begin{pmatrix}
    \mathbb{C} & 0 \\
    0 & \sigma^2_{\nbargal} \\
  \end{pmatrix},\label{eq:Sigma_diff}
\end{align}
where $\sigma^2_{\nbargal}$ has the same meaning as that of eq. \ref{eq:Sigma}.
We plug $\Sigma_{\text{diff}}$ into eq.~\ref{eq:chi2_fit} and obtain the final HOD fitting parameters. Note that the two-point clustering can be the correlation function or power spectrum. We have tested that using the power spectrum monopole and quadrupole with the fitting range $0.02\hMpc \leq k \leq 0.5\hMpc$ and the step size $0.02\hMpc$ gives reliable HOD fitting parameters. We show the best-fit $\fastpm$ HOD parameters for the $\fastpm$ LRGs, ELGs, and QSOs in table \ref{tab:hodfit}.

\subsection{CARPool method} \label{sec:carpool_method}
Given a limited number of $N$-body simulations, the measured clustering signal has large sample variance. The CARPool method, short for Convergence Acceleration by Regression and Pooling, firstly proposed by \cite{Chartier2021}, is a straightforward way to mitigate the sample variance. It utilizes the principle of control variates. Starting from the scalar case, supposing $y$ is an observable from an $N$-body simulation, we can construct a new observable $x$ as
\begin{align}
    x = y - \beta(c -\mu_c), \label{eq:carpool_scalar}
\end{align}
where $c$ is the same observable from a paired fast simulation or surrogate using the same IC as the $N$-body simulation, $\beta$ is a coefficient, and $\mu_c$ is the expected mean of $c$. Note that $\mu_c$ can be modeled by theory or estimated from a number of simulations with random ICs. Since the ICs of $c$ and $\mu_c$ are different, there will be no cross-correlation between $c$ and $\mu_c$.
Based on eq. \ref{eq:carpool_scalar}, the expectation of $x$ is unbiased from that of $y$, i.e. $\langle x\rangle=\langle y \rangle$, since $\langle c \rangle=\mu_c$. 
In addition, the sample variance of $x$ is 
\begin{align}
\sigma^2_x = \sigma^2_y - 2\beta \text{cov}(y, c) + \beta^2 \sigma^2_c + \beta^2 \sigma^2_{\mu_c}, \label{eq:sigma2_x}
\end{align}
where $\text{cov}(y,c)$ is the covariance between $y$ and $c$.
We aim to make $\sigma^2_x<\sigma^2_y$ from CARPool. By varying the coefficient $\beta$ to have $\frac{\partial \sigma^2_x}{\partial \beta}=0$, we can minimize $\sigma^2_x$. So we get
\begin{align}
    \beta = \frac{\text{cov}(y,c)}{\sigma^2_c + \sigma^2_{\mu_c}} \simeq \frac{\text{cov}(y, c)}{\sigma_c^2}. \label{eq:beta_scalar}
\end{align}
The second equation holds if $\mu_c$ is from theoretical prediction (i.e. $\sigma_{\mu_c}=0$) or from a large number of surrogates (i.e. $\sigma_{\mu_c}\ll \sigma_c$). Substituting eq.~\ref{eq:beta_scalar} into eq.~\ref{eq:sigma2_x}, we have
\begin{align}
\frac{\sigma_x^2}{\sigma_y^2} = 1 + \rho^2_{y,c}\bigg(\frac{\sigma^2_{\mu_c}}{\sigma^2_c}-1\bigg), 
\end{align}
where $\rho_{y,c}=\text{cov}(y,c)/(\sigma_y \sigma_c)$ is the Pearson correlation coefficient between $y$ and $c$. The larger cross-correlation between $y$ and $c$ (as $\rho_{y,c}\rightarrow 1$) and smaller $\sigma^2_{\mu_c}$, the smaller the sample variance of $x$ will be. In addition, we can derive the ratio of the variance of the mean $x$ and $y$ over $N$ realizations, i.e.
\begin{align}
    \frac{\sigma_{\overline{x}}^2}{\sigma_{\overline{y}}^2} = 1 + \rho^2_{y,c}\bigg(N\frac{\sigma^2_{\mu_c}}{\sigma^2_c}-1\bigg),\label{eq:sigma2_x_mean}
\end{align}
where we assume that the same $\mu_c$ is used for $N$ paired simulations, hence, averaging $x$ over $N$ realizations does not reduce the sample variance from $\mu_c$, which is counted by the factor $N$ in eq.~\ref{eq:sigma2_x_mean}. In this study, we mainly estimate $\mu_c$ from 313 $\fastpm$ simulations with random ICs. In section \ref{sec:result}, we show the sample variance suppression for one mock and $N=25$ mocks, respectively.

We can extend eq.~(\ref{eq:carpool_scalar}) and (\ref{eq:beta_scalar}) to the case of vectors, e.g. $Y=(y_1, y_2, \cdots, y_j)$, where the subscript $j$ denotes the bin  index. Assuming $\mu_C$ is known, we can calculate the covariance matrix $\Sigma_{XX}$ of a vector $X$, i.e.
\begin{align}
    \Sigma_{XX}(\beta) = \Sigma_{YY} - \beta \Sigma_{YC}^T - \Sigma_{YC}\beta^T + \beta \Sigma_{CC}\beta^T, \label{eq:cov_X}
\end{align}
where $\Sigma_{YY}$ and $\Sigma_{CC}$ are the covariance matrices of $Y$ and $C$, respectively. $\Sigma_{YC}$ is the cross covariance between $Y$ and $C$, and the superscript $T$ denotes the transpose. To minimize the determinant of $\Sigma_{XX}$, we can derive 
\begin{align}
    \beta = \Sigma_{YC} \Sigma_{CC}^{-1}.
\end{align}
If the number of the surrogates paired with $N$-body simulations is less compared to the length of vector $C$, we can not directly calculate the inverse of $\Sigma_{CC}$.\footnote{In our case, we only have 25 $\abacus$ realizations. The number of simulations is less than the number of coordinate bins of the clustering signal that we study.} We may use the singular value decomposition to do pseudo-inverse, however, the estimated $\beta$ can be unstable and reduces the CARPool performance \citep{Chartier2021,Ding2022}.
We simply set the off-diagonal terms of $\beta$ to zero following the suggestion in \cite{Chartier2021}, i.e.
\begin{align}
    \beta^{\text{diag}} = 
% \begin{pmatrix}
% \text{cov}(y_1, c_1)/\sigma_{c_1}^2 & 0 & \cdots\\
% 0 & \text{cov}(y_2, c_2)/\sigma_{c_2}^2 & \cdots\\
% \end{pmatrix}.
\begin{pmatrix}
    \text{cov}(y_1, c_1)/\sigma_{c_1}^2 & 0 & \dots & 0 \\
    0 & \text{cov}(y_2, c_2)/\sigma_{c_2}^2 & \dots & 0 \\
    \vdots & \vdots & \ddots & \vdots \\
    0 & 0 & \dots & \text{cov}(y_j, c_j)/\sigma_{c_j}^2
  \end{pmatrix}\label{eq:beta_diag}
\end{align}
Such setting is named as the univariate CARPool, which considers each vector element independent from each other. Basically, we apply the scalar case of $\beta$ for each element. In our study, we adopt $\beta^{\text{diag}}$, and estimate $\Sigma_{YC}$ and $\Sigma_{CC}$ from 25 paired $\abacus$ and $\fastpm$ realizations.

%% file: galaxy_clustering.tex
\section{Galaxy clustering statistics}\label{sec:clustering}

\subsection{Two-point correlation function}
For cubic mocks, the two-point correlation function can be calculated from \citep{Peebles1974}
\begin{align}
    \xi(s, \mu) = \frac{DD(s, \mu) - RR(s, \mu)}{RR(s, \mu)},
\end{align}
where DD and RR denote the galaxy-galaxy and random-random pairs in a given $(s,\mu)$ bin, respectively. $s$ is the modulus of the separation vector $\vb*{s}$ of a galaxy pair, i.e. $s=\sqrt{s^2_{\parallel} + s^2_{\perp}}$, and $\mu$ is the cosine angle between $\vb*{s}$ and the LoS, i.e. $\mu=s_{\parallel}/s$. We consider the RSD effect on the coordinates along the LoS\footnote{For a cubic box, we take $z$ axis as the fixed line of sight under the plan parallel approximation.}, i.e.
\begin{align}
    \vb*{s} = \vb*{r} + \frac{(1+z)\vb*{v}\cdot\hat{\vb*{z}}}{H(z)},
\end{align}
where $H(z)$ is the Hubble parameter.
For the galaxy catalog after the density field reconstruction that is discussed in section \ref{sec:recon}, we usually calculate the correlation function as
\begin{align}
    \xi(s,\mu)=\frac{DD(s,\mu) - 2SD(s,\mu) + SS(s,\mu)}{RR(s,\mu)},
\end{align}
where $S$ denotes the shifted random catalog.

Due to the anisotropy of $\xi(s, \mu)$ from RSD, we can expend $\xi(s,\mu)$ in the basis of the Legendre polynomials with the coefficients as the correlation function multipoles, i.e.
\begin{align}
    \xi_{\ell}(s) = \frac{2\ell +1}{2}\int_{-1}^{1} \xi(s, \mu) L_{\ell}(\mu) d\mu, \label{eq:xi_ell}
\end{align}
where $L_{\ell}$ is the Legendre polynomial at the order $\ell$. Due to the symmetry of $\xi(s, \mu)$ in terms of $\mu$, when $\ell$ is odd, $\xi_{\ell}$ vanishes. In our study, we mainly focus on the monopole ($\ell=0$) and quadrupole ($\ell=2$), which are widely used for the current galaxy survey analyses. We use \textsc{Pycorr}\footnote{https://github.com/cosmodesi/pycorr, which is a part of the DESI pipeline \textsc{cosmodesi}.}\citep{Sinha2020} to calculate the correlation function.

\subsection{Power spectrum}
The power spectrum is the Fourier transform of the two-point correlation function. We can also calculate the power spectrum from the density fluctuation directly, i.e.
\begin{align}
    \langle \delta(\vb*{k}) \delta(\vb*{k'})\rangle = \frac{(2\pi)^3}{V}\delta_D(\vb*{k}+\vb{k'}) P(\vb*{k}),
\end{align}
where $V$ is the data volume, $\delta_D$ is the Dirac delta function. Similar to eq. \ref{eq:xi_ell}, we can obtain the power spectrum multipoles $P_\ell(k)$, i.e.
\begin{align}
    P_{\ell}(k) = \frac{2\ell +1}{2}\int_{-1}^{1} P(k, \mu) L_{\ell}(\mu) d\mu.\label{eq:pk_ell}
\end{align}
We adopt \textsc{pypower}\footnote{https://github.com/cosmodesi/pypower}\citep{Hand2017} to calculate the power spectrum mulitpoles. We have removed the Poisson shot noise ($1/n_\text{g}$) when we show the power spectrum monopole.

\subsection{Bispectrum}
If the density field is exactly Gaussian distributed, the cosmological information is entirely encoded in the two-point statistics. However, the non-linear structure growth due to gravity induces NG in the later universe, hence, some cosmological information leaks into higher-order statistics. For the simplest case, we study the three-point statistics in Fourier space, i.e. the bispectrum,
\begin{align}
    \langle \delta(\vb*{k_1}) \delta(\vb*{k_2}) \delta(\vb*{k_3})\rangle = \delta_D(\vb*{k_1} + \vb*{k_2} + \vb*{k_3}) B(k_1, k_2, k_3),
\end{align}
where the wave vectors $\vb*{k_1}$, $\vb*{k_2}$ and $\vb*{k_3}$ form a triangle. We can normalize the bispectrum to obtain the reduced bispectrum, i.e.
\begin{align}
    Q(\theta) = \frac{B(k_1, k_2, k_3)}{P(k_1)P(k_2) + P(k_1)P(k_3) + P(k_2)P(k_3)},\label{eq:Qtheta}
\end{align}
where $\theta$ is the subtended angle between $\vb*{k_1}$ and $\vb*{k_2}$. We use \textsc{pylians3}\footnote{https://pylians3.readthedocs.io/en/master/index.html}\citep{Villaescusa-Navarro2020} to calculate the bispectrum under the triangle configurations with $k_2=2k_1=0.2\hMpc$, which relates to the scale of the BAO and RSD analyses.

\subsection{BAO reconstruction}\label{sec:recon}
The non-linear structure growth and RSD can smear the BAO signature and induce systematic shifts on the BAO scale. 
In order to increase the BAO S/N and to reduce the systematics, the BAO reconstruction technique was first proposed by \citep{Eisenstein2007b}. Since then, it has been widely studied in simulations (e.g. \citep{Seo2007, Mehta2011, Ding2018, Mariana2018}) and has become a standard tool in galaxy surveys (e.g. \citep{Padmanabhan2012, Wang2017, Hou2021}). In this study, we adopt the standard reconstruction scheme, which solves the linear displacement $\vb*\Psi$ based on the Zel'dovich approximation, i.e.
\begin{align}
    \nabla \cdot \vb*{\Psi} + \frac{f}{b_0} \nabla \cdot \left( (\vb{\Psi}\cdot \hat{\vb*{r}} ) \hat{\vb*{r}}\right) = -\frac{\delta_{\text{g}}(\vb*{s})}{b_0}, \label{eq:continuity_eq}
\end{align}
where $\delta_g(\vb*{s})$ is the galaxy density fluctuation in redshift space, $b_0$ is the linear galaxy bias, $f$ is the linear growth rate, and $\hat{\vb*{r}}$ is the LoS unit vector. In this study, we use the iterative fast Fourier transformation (IFFT) reconstruction, which solves eq. \ref{eq:continuity_eq} iteratively in Fourier space \citep{Burden2015}. We set $b_0=2.0,\, f=0.838$ and $b_0=1.2,\, f=0.888$ for LRGs and ELGs, respectively. Due to the low S/N ratio, we do not preform the BAO reconstruction on QSOs.

At small scales, the galaxy density field becomes non-linear, which breaks the linear continuity equation (eq. \ref{eq:continuity_eq}). We usually apply a Gaussian smoothing term on $\delta_\text{g}(k)$ in Fourier space to smooth out the non-linearity, i.e.
\begin{align}
    \tilde{\delta}_\text{g}(k) = \delta_\text{g}(k) \text{e}^{-k^2\Sigma_{\text{sm}}^2/2},
\end{align}
where $\Sigma_{\text{sm}}$ is the density smoothing scale. We adopt $\Sigma_{\text{sm}}=10\Mpch$ for both LRGs and ELGs. 
Once we get $\vb*{\Psi}$, we displace the positions of galaxies by $-\vb*{\Psi} - f({\vb*{\Psi\cdot \hat{r}}})\vb*{\hat{r}}$, and obtain the displaced density field, denoted as $\delta_{d}$. In addition, we shift a set of random particles\footnote{In a $2\Gpch$ cubic box, we construct a random catalog with the number density 5 times that of data.} by $-\vb*{\Psi} - f({\vb*{\Psi\cdot \hat{r}}})\vb*{\hat{r}}$ as that of the data, and obtain the shifted random catalog. We calculate the density contrast of the shifted random, denoted as $\delta_{\text{s}}$. The final reconstructed density field contrast is defined as 
\begin{align}
\delta_{\text{rec}} \equiv \delta_{d} - \delta_{s}.
\end{align}
Note that we choose the anisotropic reconstruction convention, denoted as RecSym, which contains the linear RSD signal in the reconstructed density field \citep{Eisenstein2007b, Seo2010,Mehta2011}. If we only shift the random by $-\vb*{\Psi}$, the reconstructed field removes most of the RSD signal, known as the isotropic reconstruction convention \citep{Padmanabhan2012, Anderson2014}, denoted as RecIso. In this study, we use \textsc{pyrecon}\footnote{https://github.com/cosmodesi/pyrecon, which is a part of the DESI pipeline.} to perform the reconstruction.

\subsection{BAO fitting model}
We can measure the BAO signal from the galaxy correlation function or power spectrum. Since CARPool reduces the sample variance of galaxy clustering signal, we study how much it can tighten the BAO constraints quantitatively. In addition, we can compare our result with that of the theoretical control variates \citep{Hadzhiyska2023b}. In the following, we perform the BAO fitting and adopt the BAO fitting models same as those in the BOSS and eBOSS data analyses, e.g. \citep{Beutler2017, Ross2017, Neveux2020, Bautista2021}. We briefly summarize the BAO models in both configuration and Fourier spaces. We model the anisotropic power spectrum as
\begin{align}
    P(k, \mu) = b_0^2 (1 + \beta \mu^2 R )^2 F_{\text{fog}}(k, \mu, \Sigma_s) \left[ P_{\text{sm}}(k) + \left(P_\text{lin}(k)-P_{\text{sm}}(k) \right)\text{e}^{-k^2\mu^2\Sigma_{\parallel}^2 - k^2 (1-\mu^2)\Sigma_{\perp}^2} \right],
\end{align}
where $(1+ \beta \mu^2 R)^2$ is the linear Kaiser factor \citep{Kaiser1987}, $\beta=f/b_0$ (different from the $\beta$ coefficient in CARPool), and the factor $R$ relates to the BAO reconstruction, i.e. $R = 1-\exp(-k^2\Sigma^2_{\text{sm}}/2)$ for the RecIso, and $R=1$ for the RecSym and the pre-reconstruction case \citep{Seo2016}. Due to RSD in the non-linear regime, there is the ``finger-of-God" damping, which can be modeled by $F_{\text{fog}}=1/(1+k^2\mu^2\Sigma_s^2/2)^2$ with $\Sigma_s$ as a damping parameter. We decompose the linear power spectrum into two parts: one is the smooth component without the BAO signal, denoted as $P_{\text{sm}}(k)$, which can be calculated from \cite{Eisenstein_Hu_1998}, and the other is the BAO signal, i.e. $P_{\text{lin}}(k)-P_{\text{sm}}(k)$. The non-linear structure growth and RSD can damp the BAO signal. In Fourier space, it is modeled as a Gaussian damping function with two parameters $\Sigma_{\parallel}$ and $\Sigma_{\perp}$, which account for the damping along and perpendicular to LoS, respectively. 

In real observation, we do not know the true cosmology. Usually, we assume a fiducial cosmology, and convert the observed redshifts and angular positions of galaxies into physical positions, then we measure the spatial galaxy clustering. The BAO peak position in the observed clustering shifts away from the true due to different cosmologies. Based on the standard ruler test, We can have two parameters describing such shifting along and perpendicular to the LoS, respectively, i.e. 
\begin{align}
    \alpara &= \frac{H^{\text{fid}}(z)r_s^{\text{fid}}(z_d)}{H(z) r_s(z_d)}, \label{eq:alpara}\\
    \alperp &= \frac{\DA r_s^{\text{fid}}(z_d)}{D^{\text{fid}}_{\text{A}}(z) r_s(z_d)}, \label{eq:alperp}
\end{align}
where $\DA$ is the angular diameter distance as a function of redshift, $\rs$ is the comoving sound horizon scale at the end of the baryonic-drag epoch $z_d$, and the superscript fid denotes the fiducial cosmology. We have $k^{\prime}_{\parallel}=k_{\parallel}/\alpha_{\parallel}$ and $k^{\prime}_{\perp}=k_{\perp}/\alpha_{\perp}$, where $(k^{\prime}_{\parallel}, k^{\prime}_{\perp})$ and $(k_{\parallel}, k_{\perp})$ are the true and observed coordinates, respectively.
In addition, $\alpara$ and $\alperp$ are related with $\alpha$ and $\epsilon$ \citep{Padmanabhan2008, Xu2013}, i.e.
\begin{align}
    \alpha & = (\alpha_\perp^2 \alpha_\parallel)^{\frac{1}{3}}, \label{eq:alpha}\\
    1+\epsilon & = \Big(\frac{\alpha_\parallel}{\alpha_\perp}\Big)^{\frac{1}{3}},\label{eq:epsilon}
\end{align}
where $\alpha$ describes the isotropic coordinate dilation, and $\epsilon$ quantifies the anisotropic coordinate warping. In our study, we show both sets of parameters from BAO fitting.

From real observation, we directly measure the power spectrum multipoles $P_\ell(k)$ instead of $P(k,\mu)$. We need to convert the model $P(k,\mu)$ to $P_{\ell}(k)$ using eq. \ref{eq:pk_ell}, and compare it with the observed one.\footnote{To simplify the expression, we ignore the scale factor relating to the volume change due to the difference between the fiducial and true cosmologies \citep{Beutler2017}. The scale factor is highly degenerate with the broad-band shape parameters, hence, has little influence on the BAO scale parameters.} To find the best fit parameters, we minimize the $\chi^2$ value, i.e.
\begin{align}
    \chi^2 = (P^{\text{obs}}_{\ell} - P^{\text{model}}_{\ell})^T \text{C}^{-1} (P^{\text{obs}}_{\ell} - P^{\text{model}}_{\ell})
\end{align}
with
\begin{align}
    P_{\ell}^\text{model}(k) = P_{\ell}(k) + \sum_{i=i_\text{min}}^{i_\text{max}} a_{\ell i}k^i.
\end{align}
$a_{lm}k^m$ are additional polynomial terms, which can account for non-linear galaxy bias and residual systematics, and give better fit on the broad-band shape of the power spectrum. $a_{\ell i}$ are the nuisance parameters that are marginalized over for the BAO scale measurement. In this study, we set $i_\text{min}=-1$ and $i_\text{max}=4$ with 5 nuisance parameters, and the fitting range $0.02\hMpc<k<0.3\hMpc$ with $k$ width $0.005\hMpc$.
We apply the $\textsc{barry}$\footnote{https://github.com/Samreay/Barry. We adopt the $\textsc{dynesty}$ nested sampler \citep{Speagle2020} for the fitting.} package to perform the BAO fitting. A better BAO fitting model is proposed by the recent work \citep{Chen_Howlett2024}, however, the current model is sufficient for our purpose, since we mainly compare the relative difference of the statistical error of the fitted BAO scale parameters before and after CARPool applied.

In configuration space, we can obtain $\xi_\ell(s)$ from $P_{\ell}(k)$ via Hankel transform, i.e.
\begin{align}
    \xi_\ell(s) = \frac{i^{\ell}}{2\pi^2}\int_0^{\infty}k^2 j_\ell(k s) P_{\ell}(k)dk,
\end{align}
where $j_\ell$ is the Bessel function. We fit the observation by $\xi_\ell^{\text{model}}$ with some nuisance parameters, i.e. 
\begin{align}
    \xi_{\ell}^{\text{model}}(s) = \xi_{\ell}(s) + \sum_{i=i_\text{min}}^{i_\text{max}} a_{\ell i} s^i.
\end{align}
We set $i_{\text{min}}=-1$ and $i_{\text{max}}=1$ with three nuisance parameters, and the fitting range $50\Mpch<s<150\Mpch$ with bin size $4\Mpch$.  

%% file: result.tex
\section{Result}\label{sec:result}
In this section, we show the $\fastpm$ galaxy clustering from the HOD fitting, and compare it with the $\abacus$ clustering. After applying CARPool, we demonstrate the suppression on the sample variance of the galaxy clustering. As a case study, we check the constraints on the BAO scale parameters.   
\subsection{\textsc{FastPM} LRG clustering}
\begin{figure}[htbp]
\centering
\includegraphics[width=.98\textwidth]{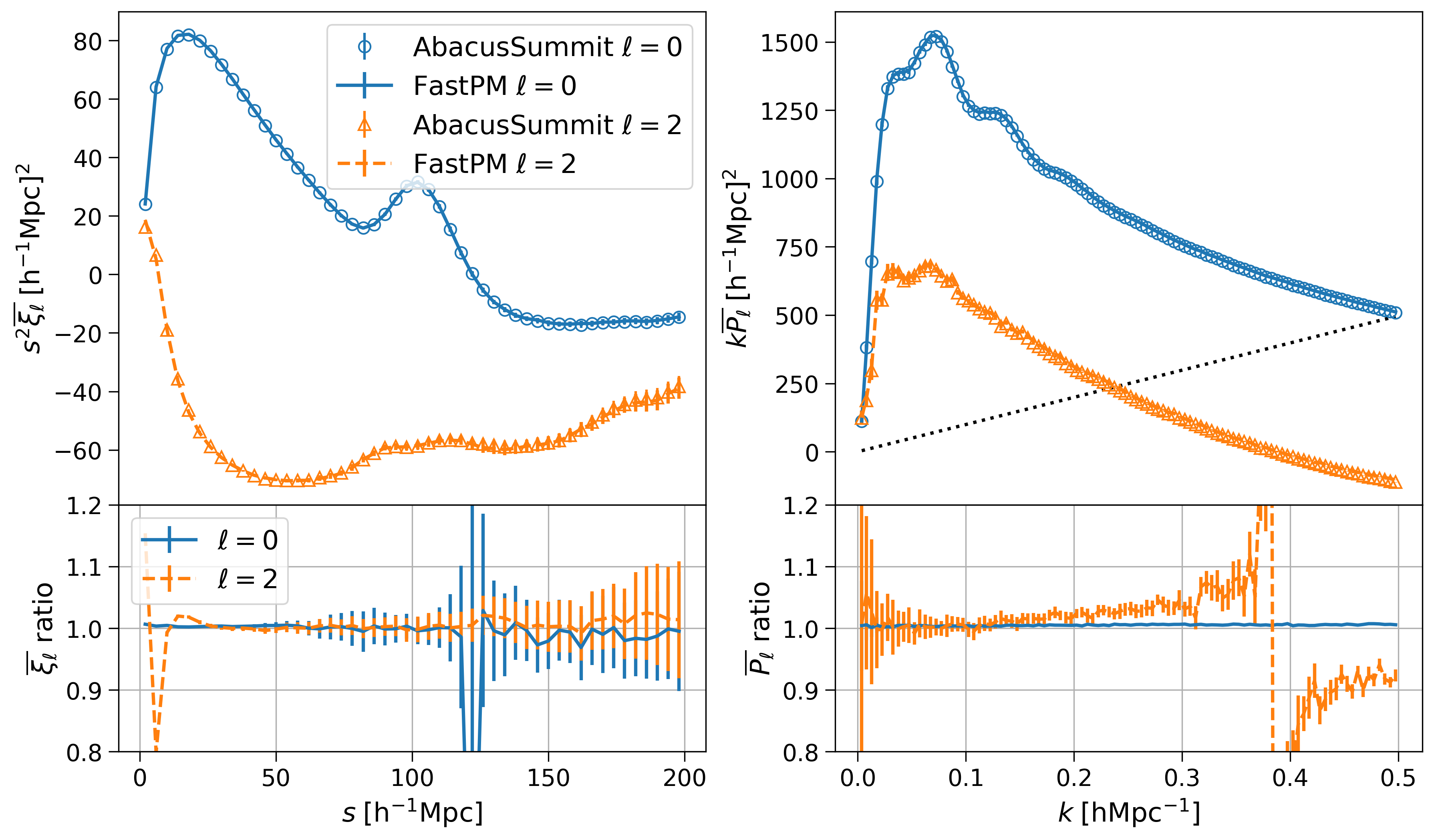}
\caption{Comparison of the mean LRG power spectrum and correlation function multipoles from 25 \textsc{AbacusSummit} and \textsc{FastPM} paired simulations. The left and right panels are for the correlation function and power spectrum, respectively. The upper panels overplot the monopoles ($\ell=0$) and quadrupoles ($\ell=2$) from the two simulations. The circular and triangular points denote the $\abacus$ monopoles and quadrupoles, respectively. Correspondingly, the solid and dashed lines show the monopoles and quadrupoles from $\fastpm$, respectively. The black dotted line denotes the mean Poisson shot noise from $\fastpm$, which closely matches to the $\abacus$ one with $0.3$ per cent difference. The lower panels display the ratios of the mean monopoles and quadrupoles between $\fastpm$ and $\abacus$. Note that some large fluctuations shown are simply due to the zero crossings of the signals. Overall, the agreement is good on the two-point clustering between the two catalogs.}\label{fig:lrg_abacus_fastpm}
\end{figure}
With the best-fit HOD parameters, we populate galaxies into $\fastpm$ halo catalogs over 25 realizations. As an example, here we display the $\fastpm$ LRG clustering.
Figure \ref{fig:lrg_abacus_fastpm} displays the comparison of the mean two-point galaxy clustering from $\abacus$ and $\fastpm$ LRG catalogs averaged over 25 realizations. The left and right panels are for the correlation function and power spectrum multipoles, respectively. The upper panels show the overall shape of the monopole ($\ell=0$) and quadrupole ($\ell=2$) moments from the two simulations. The error bars represent the standard deviation of the mean. The power spectrum monopoles shown have been subtracted by the Poisson shot noise. We denote the mean shot noise of the $\fastpm$ catalogs as the black dotted line in the upper right panel. The mean galaxy number density of $\fastpm$ matches closely to the $\abacus$ one with only $0.3$ per cent difference. The lower panels show the ratios of the monopoles and quadrupoles between $\fastpm$ and $\abacus$, respectively. 
With the HOD parameters fitted from $500\Mpch$ sub-boxes, we can match the $\fastpm$ monopoles to the $\abacus$ ones with $\sim 1$ per cent level for both Fourier and configuration spaces. In terms of the power spectrum quadrupoles, the difference is within 2 (5) per cent up-to $k=0.2\hMpc$ ($k=0.3\hMpc$), and within 10 per cent up-to $k=0.5\hMpc$. For the correlation function quadrupoles, the agreement is also within 1 per cent level except for scales $s<25\Mpch$. Note that the result we show here is conservative; we can further improve the agreement of the quadrupoles by turning HOD parameters.\footnote{We can set a narrow prior range on the HOD parameters obtained from the sub-boxes, then we do a follow-up fitting based on the original $2 \Gpch$ boxes. With larger volume and smaller sample variance, we can improve the HOD fitting at small scales. Such fine turning does not significantly enlarge the cross-correlation ($\beta$ coefficient) between the clustering from $\abacus$ and $\fastpm$, however, it may be necessary to construct a precise covariance matrix with $\fastpm$ for the galaxy clustering of Stage IV spectroscopic surveys.}
In figure \ref{fig:elg_abacus_fastpm} and \ref{fig:qso_abacus_fastpm}, we show similar results for ELGs and QSOs, respectively.

\begin{figure}
    \centering
    \includegraphics[width=0.98\textwidth]{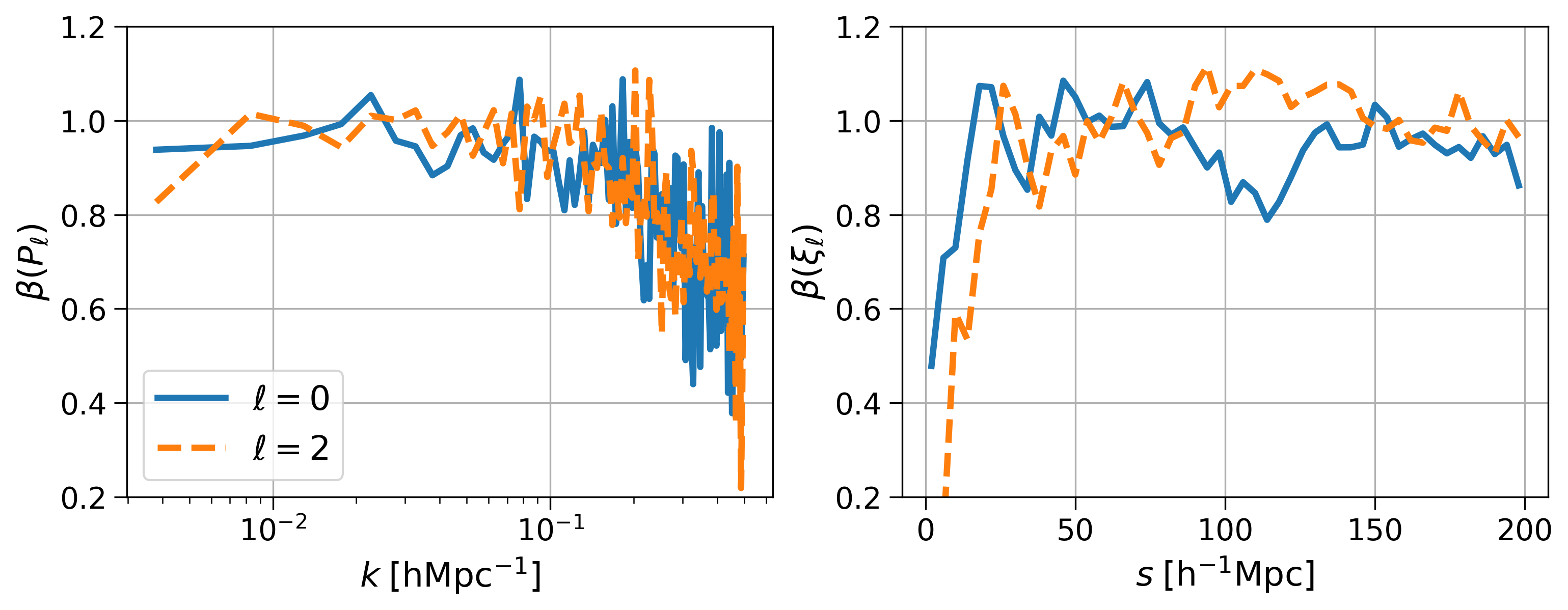}
    \caption{Diagonal terms of the $\beta^{\text{diag}}$ matrix for the LRG power spectrum and correlation function multipoles. The solid and dashed lines denote the monopole and quadrupole, respectively. $\beta$ is close to 1 at large scales, which indicates that there is high cross-correlation between the $\abacus$ and $\fastpm$ clustering.}
    \label{fig:lrg_beta}
\end{figure}
Figure \ref{fig:lrg_beta} illustrates the CARPool coefficient $\beta$ (see eq. \ref{eq:beta_diag}) for both the LRG power spectrum multipoles (in the left panel) and correlation function multipoles (in the right panel). The solid and dashed lines are for the monopoles and quadrupoles, respectively. For the power spectrum, $\beta>0.8$ at $k<0.2\hMpc$, which indicates the high cross-correlation at large scales. Correspondingly, for the correlation function, $\beta>0.8$ for $s>25\Mpch$. There is relatively large difference on $\beta$ between the correlation function monopole and quadrupole around $s=100\Mpch$, which is probably due to statistical fluctuation. Since there is high cross-correlation between the neighbouring $s$ bins of $\text{Cov}(\xi^{\text{Abacus}}_\ell,\, \xi^{\text{FastPM}}_\ell)$, statistical fluctuation on the diagonal terms from the off-diagonal terms can be large.

\begin{figure}
    \centering  \includegraphics{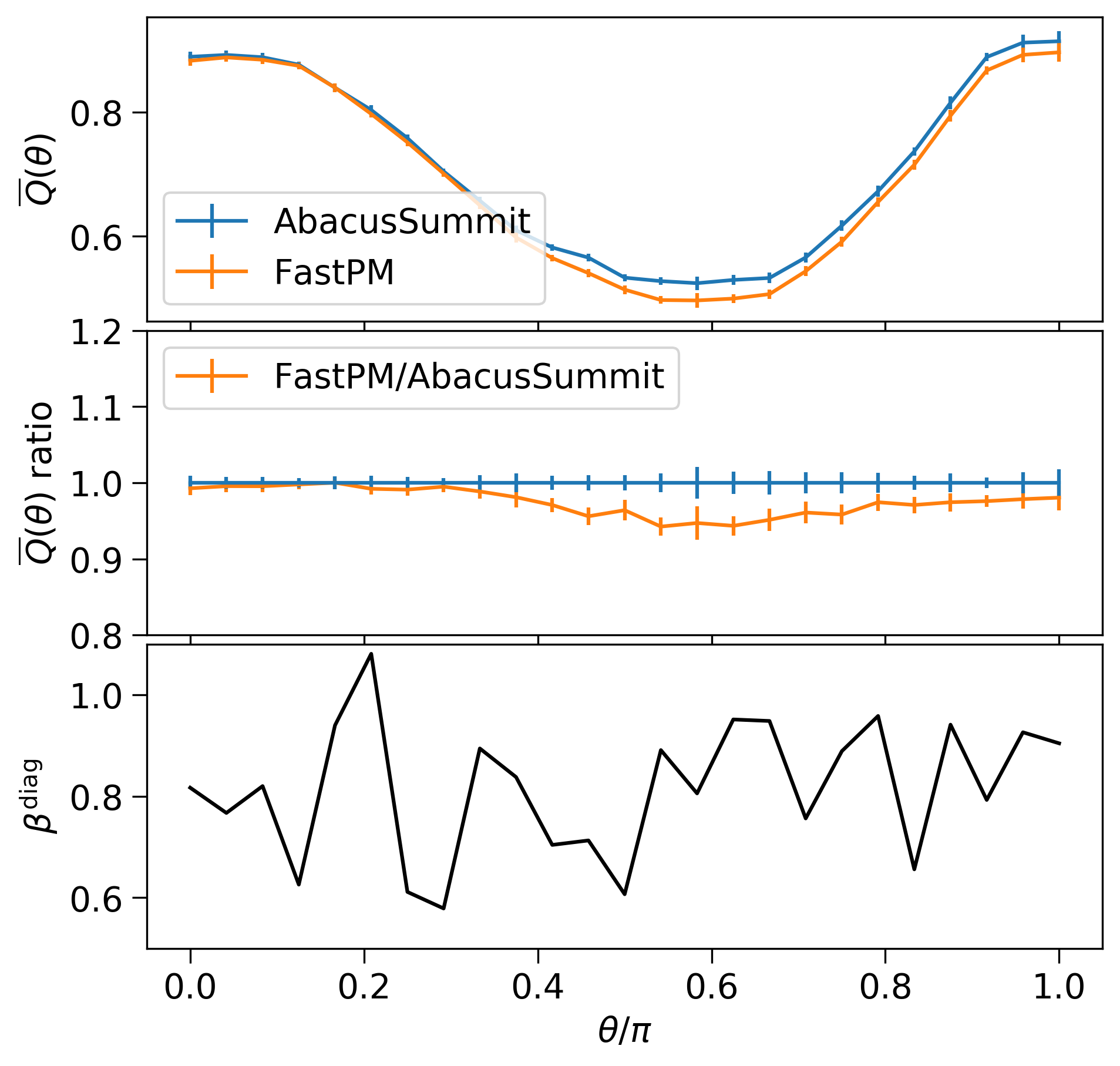}
    \caption{Upper panel: Comparison of the mean reduced bispectra $\overline{Q}(\theta)$ averaged over 25 paired \textsc{AbacusSummit} and \textsc{FastPM} LRG mocks. The triangle configurations of the bispectra are set as $k_2=2k_1=0.2\hMpc$ with $\theta$ as the subtended angle between the wave vectors $\vb*{k_1}$ and $\vb*{k_2}$. Middle panel: Ratio of $\overline{Q}(\theta)$ between $\fastpm$ and $\abacus$. Lower panel: $\beta^{\text{diag}}$ coefficient for the paired \textsc{AbacusSummit} and \textsc{FastPM} LRG mocks with $\beta\simeq 0.8$ for such configuration.}
    \label{fig:Qk_abacus_fastpm}
\end{figure}
In addition, we check the relative difference of the bispectra between the $\abacus$ and $\fastpm$ LRGs. In figure \ref{fig:Qk_abacus_fastpm}, we specifically compare the bispectra from the triangle configurations of $k_2=2k_1=0.2\hMpc$. The upper panel shows the mean reduced bispectra $\overline{Q}(\theta)$ (calculated via eq. \ref{eq:Qtheta}). The middle panel shows the ratio of $\overline{Q}(\theta)$ between $\fastpm$ and $\abacus$. The agreement is within 5 per cent, which is consistent with the finding in \cite{Variu2023}. The lower panel shows the coefficient $\beta$, which is around $0.8$, indicating the relatively high cross-correlation between the bispectra from the $\abacus$ and $\fastpm$ LRGs at these configurations. 

\subsection{Suppressing the sample variance of galaxy clustering}
In this section, we illustrate the suppression on the sample variance of galaxy clustering from CARPool.

\begin{figure}
    \centering
    \includegraphics[width=0.98\textwidth]{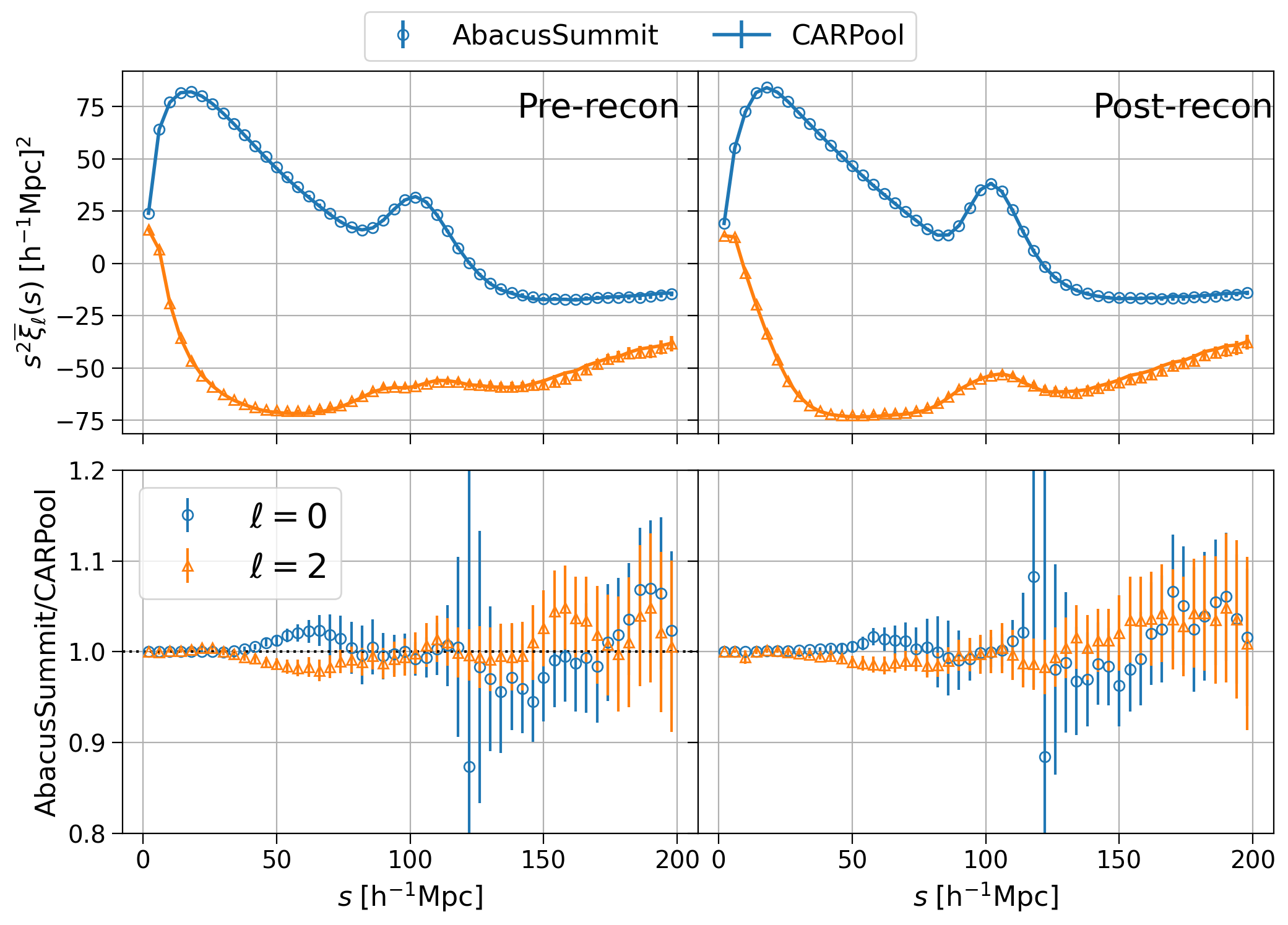}
    \caption{Upper panels: Comparison of the correlation function multipoles after CARPool applied with those from the original $\abacus$ LRG catalogs. We show the mean over 25 realizations. The markers denote the results from $\abacus$ with the circles for $\xi_0$ and the triangles for $\xi_2$. Correspondingly, the solid lines are the CARPool results. Lower panels: Ratio of the correlation function multipoles before and after CARPool applied. The left and right panels display the results before and after the BAO reconstruction, respectively. }
    \label{fig:lrg_xi02_carpool}
\end{figure}
Figure \ref{fig:lrg_xi02_carpool} compares the correlation function multipoles ($\ell=0$ and $\ell=2$) from the $\abacus$ LRGs before and after CARPool applied. The left and right panels show the results before and after the BAO reconstruction, respectively. In the upper panels, the data points represent the overall shape of the mean correlation function monopole (circular points) and quadrupole (triangular points) averaged over 25 $\abacus$ catalogs, and the lines denote the results with CARPool applied. The error bars represent $1\sigma$ error of the mean. In the lower panels, we show the ratios of the mean multipoles before and after CARPool applied. The error bars of the $\abacus$ multipoles have been rescaled by the multipoles with CARPool applied. We expect that the application of CARPool should not bias the result over a number of realizations compared to the original, as discussed in section \ref{sec:carpool_method}. We see that the ratios are close to $1.0$ with the fluctuation within $\sim 2\sigma$ level, demonstrating the unbiasedness of the CARPool result. It is true for both before and after the BAO reconstruction. Such unbiasedness is contributed from a relatively large set of $\fastpm$ mocks used to estimate the surrogate mean clustering statistics for the $\fastpm$ mocks paired with $\abacus$. Note that the large fluctuation of the monopoles at around $120\Mpch$ is simply due to the zero crossing of the signal.

\begin{figure}
    \centering
    \includegraphics[width=0.98\textwidth]{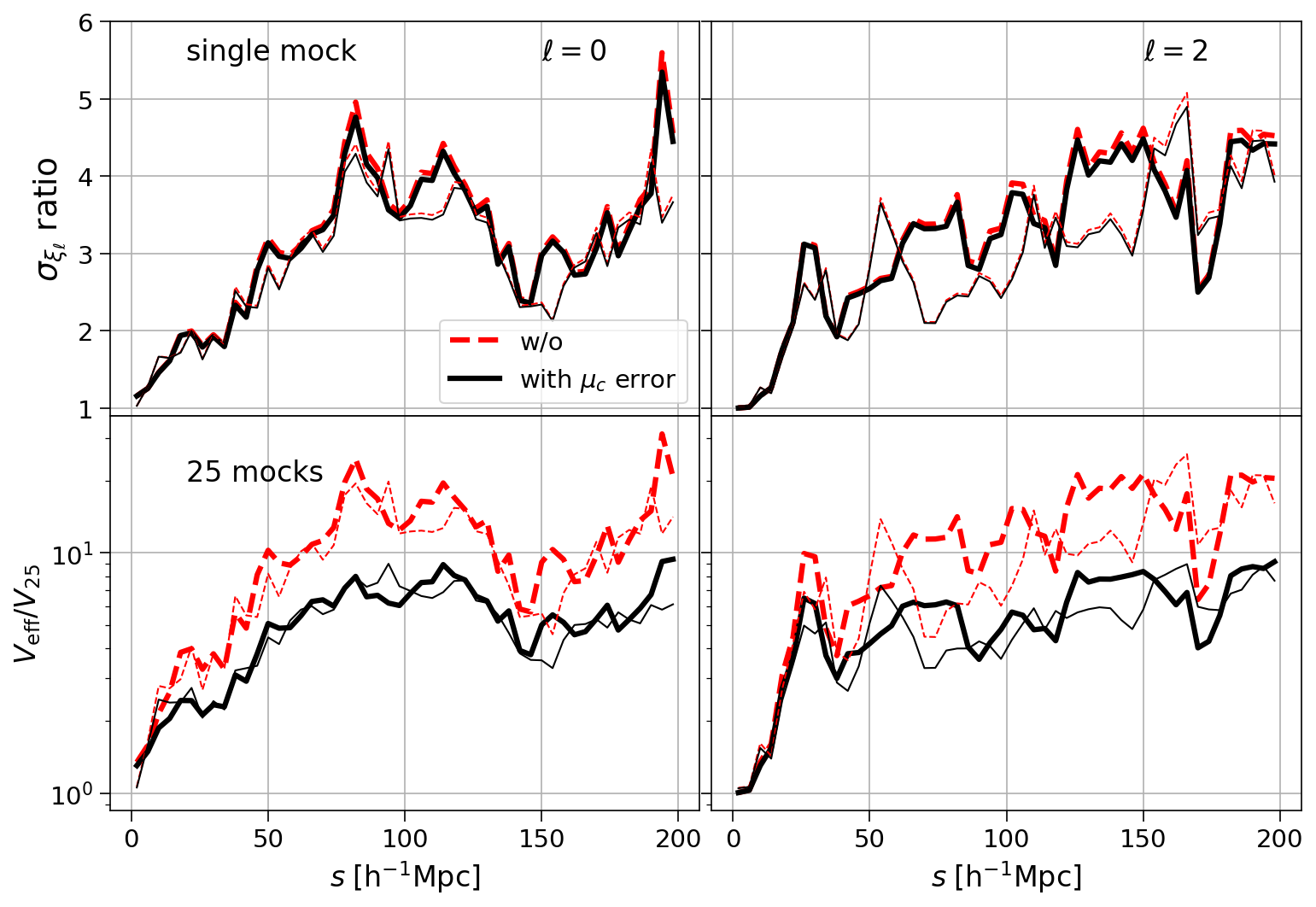}
    \caption{Upper panels: Reduction of the standard deviation of the correlation function multipoles from CARPool. We consider the standard deviation in terms of one single mock. Lower panels: Increase on the effective volume of 25 mocks from CARPool. The left and right panels are for the monopole and quadrupole, respectively. The black solid lines consider the error of $\mu_c$ in the CARPool result, while the red dashed lines do not, and represent the optimal gains. We compare the results before and after the BAO reconstruction, shown as the thick and thin lines, respectively. }
    \label{fig:err_veff_lrg}
\end{figure}
Figure \ref{fig:err_veff_lrg} illustrates the reduction on the sample variance of the $\abacus$ LRG catalogs from CARPool. The upper panels plot the ratios of the standard deviation of the $\abacus$ LRG correlation function multipoles before and after CARPool applied. We get the standard deviation from the diagonal terms of the correlation function covariance matrices. The left and right panels are for the monopole and quadrupole, respectively. With CARPool, the standard deviation of one single mock is reduced by a factor of $3\sim 4$ at the scale $50\Mpch<s<200\Mpch$. At $s<50\Mpch$, the reduction factor is less but still larger than 1, slightly better than that from the theoretical control variates, e.g. figure 7 of \cite{Hadzhiyska2023b}.
In the case that we consider the sample variance of $\mu_c$ (in eq.~\ref{eq:carpool_scalar}) estimated from the $\fastpm$ catalogs with random ICs, the result is shown as the black solid lines. Ignoring the $\mu_c$ error, we have the red dashed lines, which represent the optimal results. The difference between the black and red lines is small, indicating that the contribution of $\mu_c$ error is negligible on the sample variance in terms of one single mock. In addition, we compare the results before and after the BAO reconstruction, shown as the thick and thin lines, respectively. They have similar amplitude, indicating that the reconstruction does not affect much on the CARPool performance.

The lower panels of figure \ref{fig:err_veff_lrg} show the increase on the volume over 25 mocks, denoted as $V_{25}$. We calculate it based on eq.~\ref{eq:sigma2_x_mean}, i.e. $V_{\text{eff}}/V_{25} = \sigma^2_{\overline{y}}/\sigma^2_{\overline{x}}$. Without the sample variance of $\mu_c$, we can increase the simulation volume a factor of $\sim 10$ at $50<s<200\Mpch$ from CARPool, shown as the dashed lines. Considering the $\mu_c$ error estimated from 313 $\fastpm$ catalogs, the gain is degraded to a factor of $\sim 5$. Hence, the bottleneck of CARPool method is to estimate the surrogate mean $\mu_c$ precisely, which is limited by the number of $\fastpm$ realizations in our case. To resolve such issue, one way is to theoretically model the surrogate mean without sample variance \citep{Kokron2022,DeRose2023a,DeRose2023b,Hadzhiyska2023b}. Another way we propose is to use surrogates with the fixed-amplitude or fixed-and-paired ICs, which can effectively reduce the sample variance of surrogate mean. We have validated the performance based on the halo clustering in our previous work \citep{Ding2022}. Here we extend it to the galaxy clustering. In the appendix, figure~\ref{fig:veff_ratio_lrg} illustrates that we can reach about $\sim 80$ per cent of the optimal $V_{\text{eff}}$ in the case when we use 200 fixed-amplitude $\fastpm$ catalogs to estimate $\mu_c$. With a smaller set of fixed-amplitude $\fastpm$ realizations in CARPool, we obtain a larger effective volume compared to the one using simulations with regular ICs. We recommend such implementation for the CARPool application.

\begin{figure}
    \centering
    \includegraphics[width=0.98\linewidth]{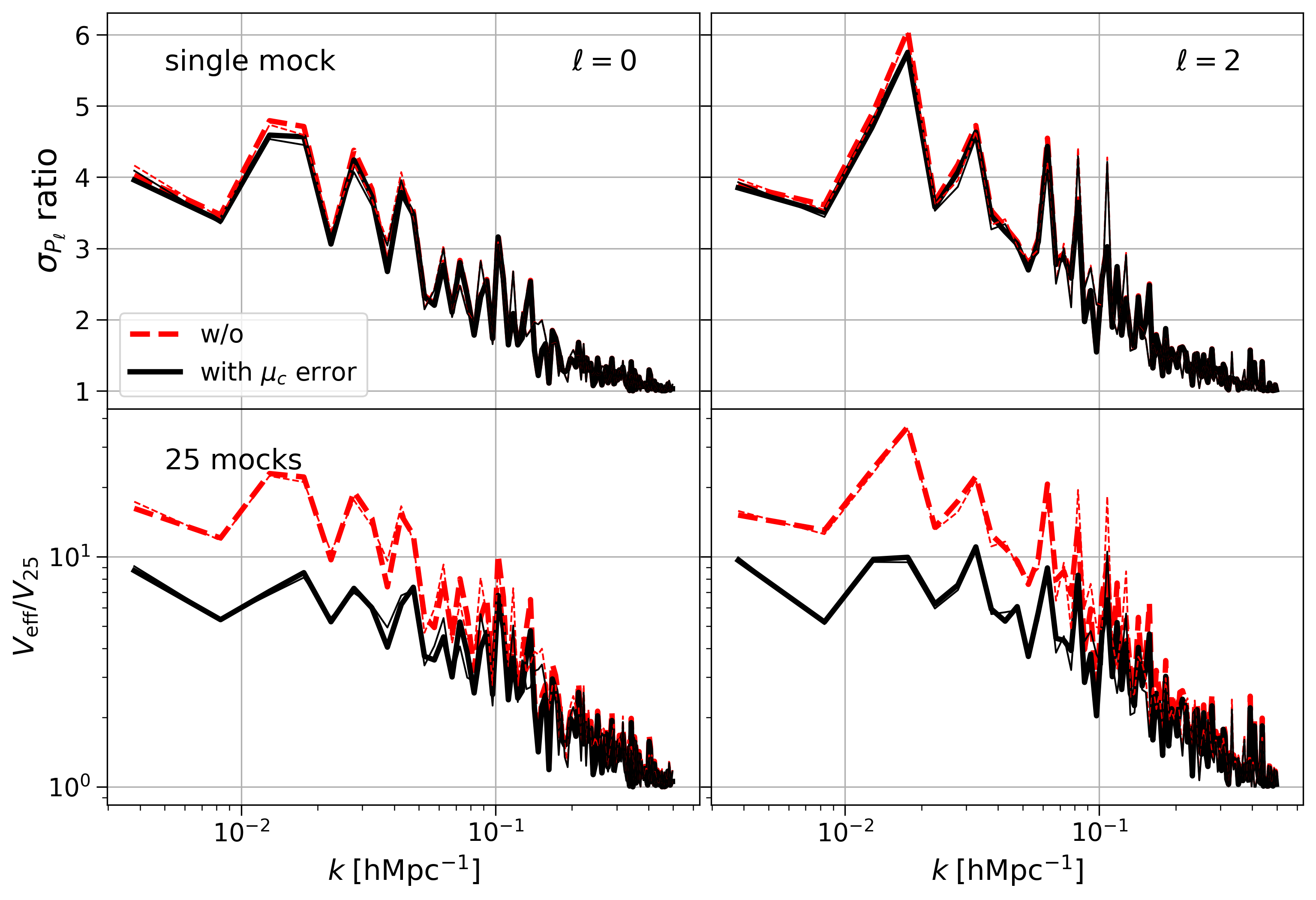}
    \caption{Similar to figure \ref{fig:err_veff_lrg} but for the $\abacus$ ELG power spectrum multipoles.}
    \label{fig:err_veff_elg}
\end{figure}

Similar to figure \ref{fig:err_veff_lrg}, figure \ref{fig:err_veff_elg} shows the results for the $\abacus$ ELGs based on the power spectrum monopoles and quadrupoles. At $k<0.1\hMpc$, the reduction on the standard deviation of one single mock is larger than a factor of $2$, and the effective volume of 25 mocks increases about $4$ times. Such significant suppression on the sample variance at large scales can be beneficial for a tighter constraint on the theoretical systematics for BAO, RSD and $\fnl$.
We notice that the $\sigma$ ratio (or $V_{\text{eff}}/V_{25}$) before and after the BAO reconstruction varies less for the power spectrum case, compared to that of the correlation function (figure \ref{fig:err_veff_lrg}). It is probably due to the larger cross-correlation between the diagonal and off-diagonal terms in the correlation function covariance matrix. 

\begin{figure*}
    \centering
    \includegraphics[width=0.98\textwidth]{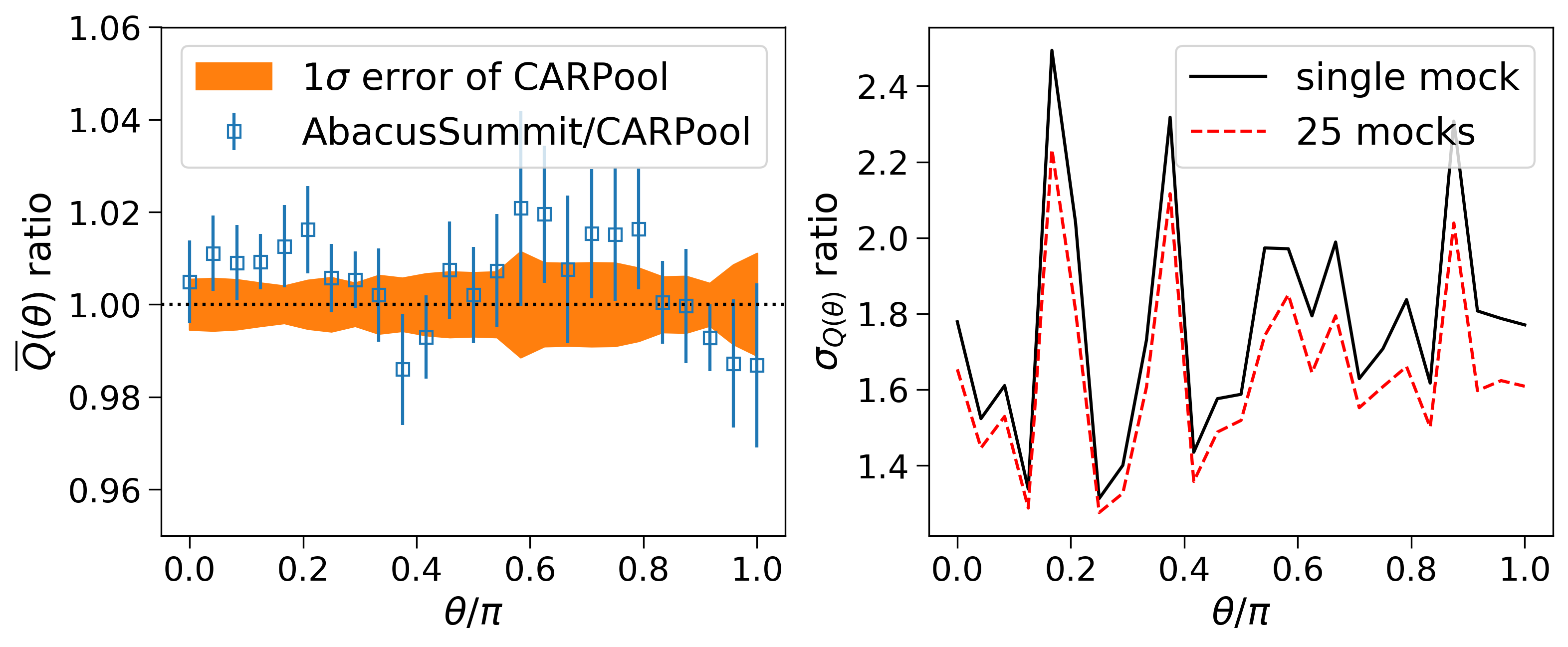}
    \caption{Left panel: Ratio of the mean $Q(\theta)$ of the $\abacus$ LRGs before and after CARPool applied shown as the blue squares. The error bars denote $1\sigma$ error of the mean, and the shaded region depicts the error bar after CARPool applied. Right panel: Ratio of the standard deviation of $Q(\theta)$ before and after CARPool. The black solid line is the case of one single mock, while the red dashed line represents for the mean of 25 mocks.}
    \label{fig:Qx}
\end{figure*}

Our previous study \cite{Ding2022} has shown that CARPool can effectively reduce the sample variance of the halo bispectrum with the triangle configurations $k_2=2k_1=0.2\hMpc$. In this study, we further check that it is true for the galaxy bispectrum as well, as shown in figure \ref{fig:Qx}. In the left panel, we check the ratio of the $\abacus$ LRG reduced bispectrum (eq. \ref{eq:Qtheta}) before and after CARPool, shown as the squares. It fluctuates around $1.0$ without noticeable bias. The orange shaded region denotes the standard deviation of the result with CARPool. In the right panel, we display the ratio of the standard deviation before and after CARPool applied. The solid line is for the case of one single mock, and the dashed line is for the mean of 25 mocks. The CARPool method reduces the standard deviation by $\sim 1.6$ times for both cases. 

\subsection{Improving the BAO constraints}
We have shown that the sample variance of galaxy clustering can be effectively suppressed by CARPool. Such clustering with smaller sample variance can be useful to constrain theoretical systematics tighter, given a limited number of $N$-body simulations. Here we perform the BAO fitting as an example. We demonstrate that fitting the galaxy clustering with CARPool applied outputs unbiased result but with smaller uncertainties on the BAO scale parameters. 
\begin{table*}
	\centering
	\begin{tabular}{|c c c c c c c c c |}
		\hline	
    & $\langle \alpha \rangle$ & $\sigma_{\alpha}$ & $\langle \epsilon \rangle$ & $\sigma_{\epsilon}$ & $\langle \alpara \rangle$ & $\sigma_{\alpara}$ & $\langle \alperp \rangle$ & $\sigma_{\alperp}$\\ \hline
    &\multicolumn{8}{c|}{Pre-recon LRGs}\\ \cline{2-9} 
		
Pre-CARPool & 1.0039 & 0.0044 & 0.0031 & 0.0073 & 1.0104 & 0.0170 & 1.0008 & 0.0071 \\
Post-CARPool & 1.0046 & 0.0026 & 0.0016 & 0.0034 & 1.0078 & 0.0086 & 1.0031 & 0.0031 \\
ratio &  & 1.6835 &  & 2.1465 &  & 1.9815 &  & 2.3018 \\
        \hline
        &\multicolumn{8}{c|}{Post-recon LRGs}\\ \cline{2-9} 
Pre-CARPool & 1.0000 & 0.0031 & -0.0003 & 0.0048 & 0.9995 & 0.0105 & 1.0003 & 0.0052 \\
Post-CARPool & 1.0007 & 0.0020 & 0.0004 & 0.0026 & 1.0015 & 0.0063 & 1.0003 & 0.0024 \\
ratio &  & 1.5449 &  & 1.8516 &  & 1.6744 &  & 2.1344 \\
        \hline
        &\multicolumn{8}{c|}{Pre-recon ELGs}\\ \cline{2-9} 
Pre-CARPool & 1.0031 & 0.0053 & 0.0022 & 0.0057 & 1.0075 & 0.0136 & 1.0010 & 0.0070 \\
Post-CARPool & 1.0030 & 0.0038 & 0.0032 & 0.0028 & 1.0094 & 0.0072 & 0.9998 & 0.0043 \\
ratio &  & 1.3958 &  & 2.0717 &  & 1.8693 &  & 1.6276 \\
        \hline
        &\multicolumn{8}{c|}{Post-recon ELGs}\\ \cline{2-9} 
Pre-CARPool & 1.0015 & 0.0036 & -0.0013 & 0.0026 & 0.9989 & 0.0066 & 1.0028 & 0.0042 \\
Post-CARPool & 1.0020 & 0.0027 & -0.0002 & 0.0022 & 1.0017 & 0.0040 & 1.0022 & 0.0042 \\
ratio &  & 1.3317 &  & 1.1627 &  & 1.6616 &  & 1.0001 \\
        \hline
	\end{tabular}
 \caption{Summary of the BAO scale parameters fitted from the $\abacus$ LRGs and ELGs two-point clustering before and after CARPool applied, denoted as Pre-CARPool and Post-CARPool, respectively. For each tracer, we also compare the results before and after the BAO reconstruction, denoted as Pre-recon and Post-recon, respectively.}\label{tab:baofit_lrg_elg}
\end{table*}

\begin{figure}
    \centering
    \includegraphics[width=0.49\linewidth]{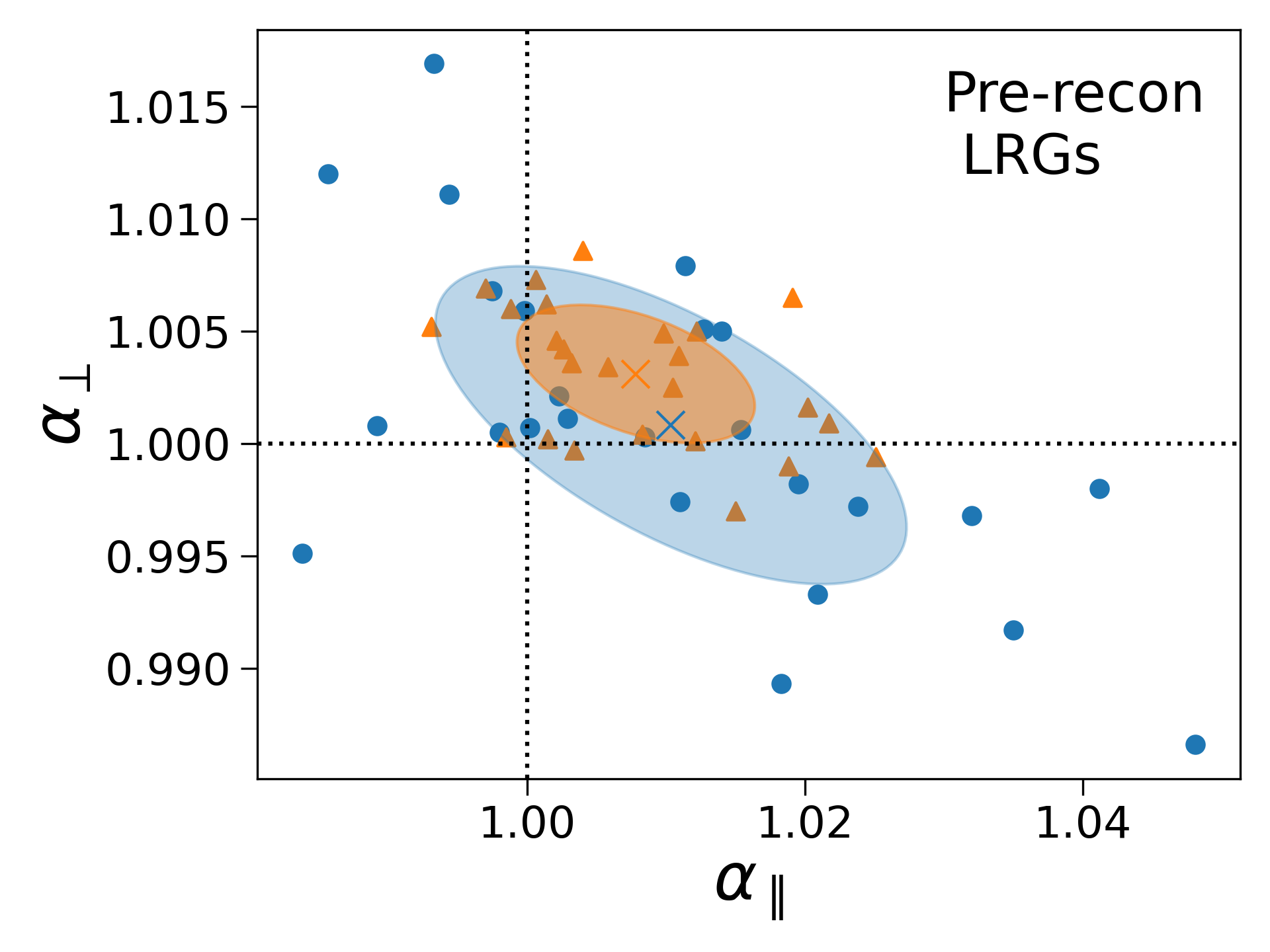}
    \includegraphics[width=0.49\linewidth]{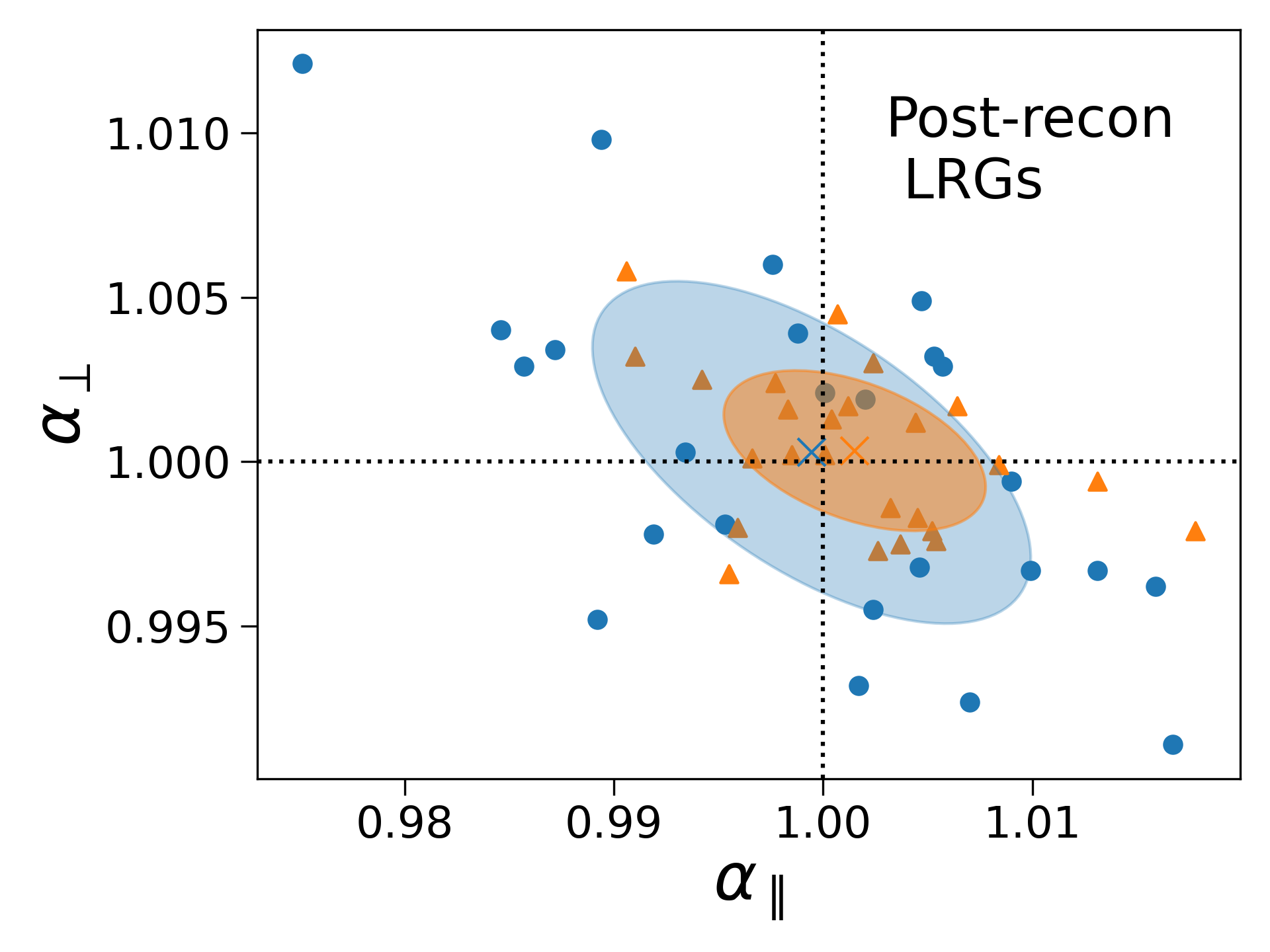}
    \includegraphics[width=0.49\linewidth]
    {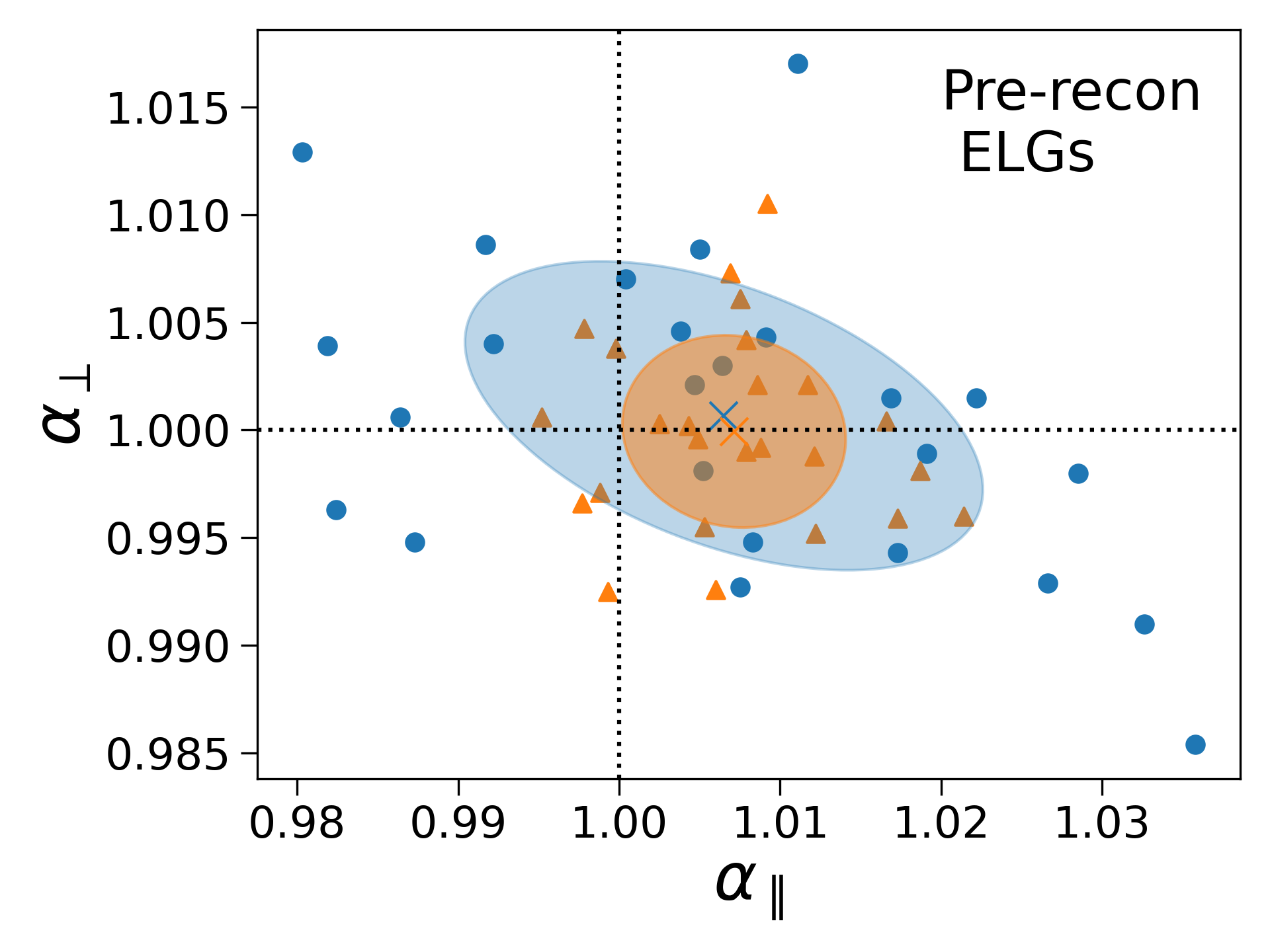}
    \includegraphics[width=0.49\linewidth]{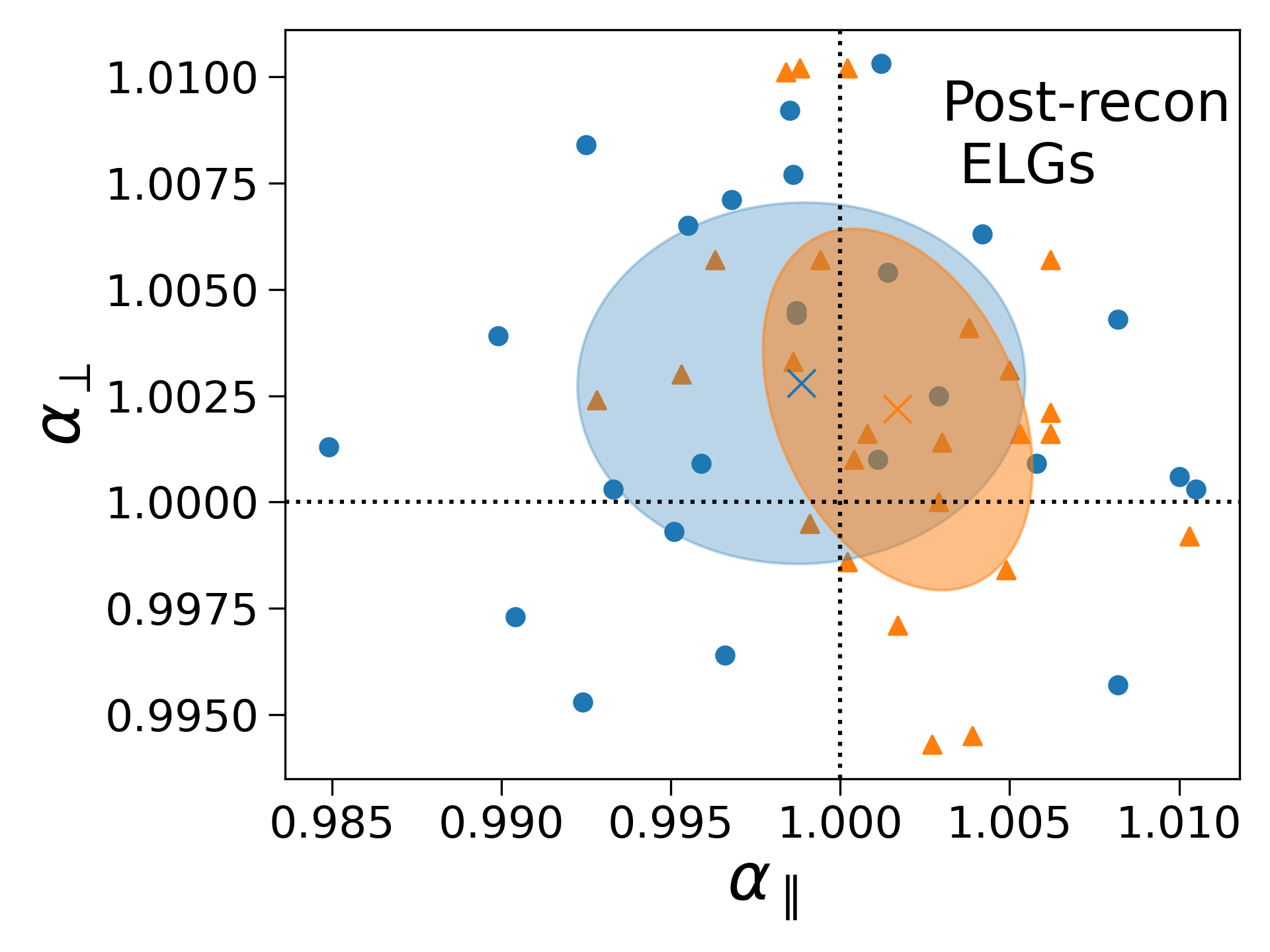}
    \caption{BAO fitting parameters $\alpara$ and $\alperp$ fitted from the \textsc{AbacusSummit} LRGs and ELGs. The left and right panels show the results before and after the BAO reconstruction, respectively. For LRGs, we fit the correlation function monopole and quadrupole in the range $50\Mpch<s<150\Mpch$. For ELGs, we fit the power spectrum monopole and quadrupole in the range $0.02\hMpc<k<0.3\hMpc$. Each data point represents the fitting result from one mock. The blue and orange colors denote the results before and after CARPool applied, respectively. For each case, we plot a contour based on the mean (denoted by the crosses), the standard deviation, and the cross-correlation coefficient between $\alpara$ and $\alperp$ over 25 realizations. With CARPool, the contour shrinks significantly, indicating the suppression on the sample variance of the BAO scale parameters over realizations.}
    \label{fig:alphas_lrg_elg}
\end{figure}
In the fitting with \textsc{barry}, we set $\alpha$, $\epsilon$, $b_0$, $\beta$, $\Sigma_{\parallel}$,  $\Sigma_{\perp}$, and $\Sigma_s$ as free parameters. We can convert $\alpha$ and $\epsilon$ to get $\alpara$ and $\alperp$ based on eq. \ref{eq:alpara} and \ref{eq:epsilon}. We get the mean $\alpha$ and $\epsilon$ (or $\alpara$ and $\alperp$) after marginalizing over the other parameters including the nuisance parameters.
In figure \ref{fig:alphas_lrg_elg}, the upper panels show the marginalized mean of $\alpara$ and $\alperp$ from fitting the correlation function monopoles and quadrupoles over 25 $\abacus$ LRG mocks. 
The left and right panels are the results before and after the BAO reconstruction, respectively. When we fit the clustering with and without CARPool applied, we use the same covariance matrix, which is estimated from 313 $\fastpm$ catalogs with the random ICs. Strictly speaking, the covariance matrix of the galaxy clustering after CARPool should be different from the one before CARPool (as shown in eq. \ref{eq:cov_X}). Here we only focus on the potential application of the CARPool method, and leave more details for future work.
For the $\abacus$ clustering with reconstruction, we apply the same reconstruction process on the $\fastpm$ catalogs to obtain the post-reconstruction covariance matrix.
In each panel, the blue circles and the orange triangles represent the results before and after CARPool applied, respectively. Each data point is fitted from one mock. 
To visually compare the scattering of the data points between the two cases, we plot two shaded contours, whose size and ellipticity are based on the standard deviations of $\alpara$ and $\alperp$, as well as their cross-correlation coefficients. We match the colors of contours to those of data points. The mean values of $\alpara$ and $\alperp$ are denoted by the crosses, which are close to each other from the two cases, indicating that there is little bias on the fitted parameters using the clustering with CARPool. 
Since the BAO reconstruction removes most of systematics from the non-linear structure growth and RSD, the resultant density field becomes more linear, and the BAO scale parameters are closer to 1. 
As shown in the right panel, the mean values of $\alpara$ and $\alperp$ (shown as the crosses) are close to 1 for both cases before and after CARPool applied. 

Similarly, we show the fitting results of the $\abacus$ ELGs in the lower panels of figure \ref{fig:alphas_lrg_elg}. We fit the power spectrum monopoles and quadrupoles over 25 realizations. The results are similar to those of LRGs, illustrating that we can apply galaxy clustering with CARPool in both configuration and Fourier spaces for systematics study. We notice that there is a larger bias ($\sim 0.25\%$) on the mean $\alperp$ of post-reconstruction for both cases before and after CARPool applied compared to that of LRGs, indicating some systematic error. We have tested changing the density smoothing parameter from $10\Mpch$ to $15\Mpch$, which is a parameter can be optimized \citep{KP4s3-Chen, KP4s4-Paillas} for the BAO reconstruction, however, it reduce the bias little. Increasing the random sample size from 5 to 20 times of the data size does not help neither, when we perform reconstruction for the $\abacus$ ELGs. We suspect that the systematics can be caused by the covariance matrix from the limited number of $\fastpm$ realizations. If we replace our covariance matrix by the one from 1000 realizations of \textsc{EZmocks}\citep{Chuang2015}, the bias can be largely mitigated. We discuss the result in appendix \ref{sec:ezmock_cov}. In terms of the dispersions of the BAO fitting parameters among 25 $\abacus$ mocks, we have checked that our results agree well with the ones from the studies of the LRG and ELG HOD systematics \citep{KP4s10-Mena-Fernandez,KP4s11-Garcia-Quintero}. In their studies, the ``FirstGen'' mocks are the same as our $\abacus$ mocks.

Table \ref{tab:baofit_lrg_elg} lists the quantitative values of the BAO fitting parameters for the $\abacus$ LRGs and ELGs, respectively. In the first few columns, we show the mean of $\alpha$ and $\epsilon$ over 25 mocks, as well as their standard deviation, i.e. $\sigma_{\alpha}$ and $\sigma_{\epsilon}$. In the last few columns, we show the mean and standard deviation of $\alpara$ and $\alperp$, respectively. In addition, we display the ratio of the mean and standard deviation before and after CARPool applied (denoted as Pre-CARPool and Post-CARPool). For the mean values of the parameters, we expect that their ratios are close to 1, which is the case except for $\epsilon$. It is simply due to numerical fluctuation, since $\epsilon$ is close to 0 especially for the post-reconstruction. If we take the difference between $\epsilon$ from Pre- and Post-CARPool, it is less than or comparable with the standard deviation of the mean, i.e. $\sigma_{\epsilon}/\sqrt{25}$. For the ratio of the standard deviation before and after CARPool, we see that it is larger than 1. For the LRGs after the reconstruction, the constraints on $\alpha$, $\epsilon$, $\alpara$ and $\alperp$ are enhanced from CARPool by a factor of 1.54, 1.85, 1.67, and 2.13, respectively. For the ELGs after the reconstruction, the improved factors are 1.33, 1.16, 1.66, and 1.0, respectively. Our result is generally consistent with that from the theoretical control variates \cite{Hadzhiyska2023b, KP4s10-Mena-Fernandez, KP4s11-Garcia-Quintero}, the relative difference can be caused by different settings on the reconstruction and BAO fitting.\footnote{For example, \cite{Hadzhiyska2023b} fit the reconstructed ELG correlation function multipoles with the pre-reconstructed covariance matrix, which is based on \textsc{EZmocks}. The reduction factors for the dispersions of $\alpha$, $\epsilon$, $\alpara$ and $\alperp$ are 1.5, 1.5, 1.8, 1.2, respectively. If we use their input but replace the correlation function multipoles by the power spectrum multipoles, we get the reduction factors equal to 1.33, 1.52, 1.74, 1.13, closer to theirs.}

%% file: conclusions.tex
\section{Conclusions and discussions}\label{sec:conclusion}
To study the CARPool performance on galaxy clustering, we utilize the $\fastpm$ simulations, which use the same cosmology as the $\abacus$ simulations.
Based on the HOD models, we produce the $\fastpm$ galaxy catalogs, which can mimic the DESI-like LRGs, ELGs, and QSOs from $\abacus$. For each tracer, the $\fastpm$ galaxy number density and the two-point clustering are well matched to those of $\abacus$.
The agreement is within a few per cent or better for both the correlation function and power spectrum multipoles at the scales relating to the BAO and RSD analyses. Although the higher-order galaxy clustering is not included in the HOD fitting, the resultant $\fastpm$ galaxy bispectrum agrees well with the $\abacus$ one. For LRGs, the agreement is better than $5$ per cent at the triangle configurations with $k_2=2k_1=0.2\hMpc$.  

In addition, we utilize the paired $\fastpm$ and $\abacus$ simulations with the same ICs. There is high cross-correlation between the galaxy clustering from the two sets of simulations, which encourages the application of CARPool. In our previous study, we have shown that CARPool can effectively reduce the sample variance of $\abacus$ halo clustering \citep{Ding2022}. In this work, we adopt the same methodology but extend it to the galaxy clustering. We have examined in detail that the galaxy clustering with CARPool applied is unbiased. We obtain the quantitative amount of the sample variance reduction from CARPool. 
Specifically, for the $\abacus$ LRG correlation function multipoles, we can reduce the standard deviation of a single mock by a factor of $3\sim 4$ at the scale range $50\Mpch<s<200\Mpch$. The effective volume of 25 mocks can be increased at least by a factor of 5 given the current number of $\fastpm$ simulations with random ICs. We expect that the optimal performance is a factor of $\sim 10$. It can be largely achieved with the assistance of the fixed-amplitude $\fastpm$ simulations, which can significantly reduce the sample variance of the ensemble average of the statistics, i.e. $\mu_c$ in eq.~(\ref{eq:carpool_scalar}). For the power spectra of ELGs, we find that CARPool can increase the effective volume larger than 4 times at $k<0.1\hMpc$. We have checked the CARPool performance for the galaxy clustering after the BAO reconstruction applied. We show that there is no obvious difference on the cross-correlation coefficient (i.e. $\beta$) between the $\abacus$ and $\fastpm$ galaxy clustering before and after the BAO reconstruction.

For QSOs, we find that the cross-correlation coefficient of the two-point clustering is less than 1.0 between $\fastpm$ and $\abacus$ at large scales. We notice that the work \cite{DeRose2023a} based on the Zel'dovich control variates has found similar result (see their figure 8). It is probably due to the high shot noise over all relevant scales that degrades the cross-correlation.  

As a case of application, we study the improvement on the BAO constraints from the galaxy clustering with CARPool applied. We perform the BAO fitting on the $\abacus$ two-point galaxy clustering. For LRGs, we fit the correlation function monopole and quadrupole from each realization. We study the mean and standard deviation of the fitted BAO scale shifting parameters along and perpendicular to the LoS, i.e. $\alpara$ and $\alperp$, which are related with the isotropic and anisotropic parameters, i.e. $\alpha$ and $\epsilon$. The mean of the fitted BAO scale parameters are close to each other before and after CARPool, illustrating that there is no bias from CARPool. The standard deviation is significantly reduced after CARPool applied. The reduction factors for $\alpha$, $\epsilon$, $\alpara$ and $\alperp$ are respectively 1.54, 1.85, 1.67, and 2.13 for the case after the BAO reconstruction. For ELGs, we perform the BAO fitting similarly but on the power spectrum monopoles and quadrupoles. The improved factors of the constraints on $\alpha$, $\epsilon$, $\alpara$ and $\alperp$ are 1.33, 1.16, 1.66, and 1.0, respectively.

Since the suppression on the sample variance is mostly significant at large scales, we expect that our work can be useful for tighter constraints on the theoretical systematics of the RSD and $\fnl$ models as well. With some observational systematics, such as the imaging weight and fiber assignment, added on the galaxy catalogs, we are able to study their impacts on the clustering signal and the fitted cosmological parameters more precisely with the assistance of CARPool. We leave such study in future work. 

Compared with the method of theoretical control variates, the main caveat of CARPool is computational time consuming. It takes tens of millions of CPU hours to generate hundreds of fast simulations in our case. To mitigate such issue, we may choose some cheaper simulations, such as \textsc{EZmock} \citep{Chuang2015, ZhaoCheng2021}. On the other hand, we usually run a large number of fast simulations to estimate covariance matrices for galaxy clustering analysis. If these fast simulations are ready, it can be a byproduct to perform CARPool. 
Although It is relatively cheap to fit HOD parameters for $\fastpm$ galaxies, performing density field reconstruction and calculating two-point statistics can take some amount of computational time for hundreds of galaxy mocks. Again, these products are not only used by CARPool, but can be useful for other purposes. In addition, it is straightforward to apply CARPool on higher order statistics, such as bispectrum, compared to the current stage of theoretical control variates.

%% file: appendix.tex
\appendix

\section{$\fastpm$ HOD parameters}
\begin{table}[htbp]
\small
\centering
\resizebox{\columnwidth}{!}{%
\begin{tabular}{|c | c | c| c | c | c | c |}
\hline
\hline
& \multicolumn{2}{c|}{LRGs} & \multicolumn{2}{c|}{ELGs} & \multicolumn{2}{c|}{QSOs} \\
\hline
Parameters & prior  & best-fit & prior  & best-fit & prior  & best-fit \\
\hline
log$M_{\text{cut}}$      & (11.6, 13.6) & 12.69 [12.687] & (10.7, 13.0) & 11.03 [11.22] & (11.6, 13.6) & 13.25 [12.473]\\ \hline
log$M_1$                 & (9.0, 14.0)  & 11.71 [13.71] & (12.0, 15.0) & 14.38 [12.28] & (9.0, 14.0)  & 11.88 [14.0]\\ \hline
$\sigma_{\text{log}M_h}$ & (0.0, 4.0)   & 0.44  [0.5] & (0.0, 5.0)   & 1.12 [0.6] & (0.0, 4.0)   & 1.71 [1.0]\\ \hline
$\kappa$                 & (0.0, 20.0)  & 4.08 [1.0]  & (0.5, 2.0)   & 2.32 [0.01] & (0.0, 20.0)  & 3.56 [1.0]\\ \hline
$\alpha$                 & (0.0, 1.3)   & 0.12 [0.8]  & (0.0, 5.0)   & 0.49 [0.007] & (0.0, 1.3)   & 0.0455 [1.0]\\ \hline
$v_{\text{disp}}$        & (0.7, 1.5)   & 1.01   & (0.8, 1.6)   & 1.32  & (0.7, 1.5)   & 0.58  \\ \hline
$\gamma$                 &              &        & (5.0, 10.0)  & 6.28 [4.7] &              & \\ \hline
$A$                      &              &        & (0.0, 1.0)   & 0.14  &              & \\ \hline
$p_{\text{max}}$         &              &        &              &    [0.65]   & (0.0, 1.0)   & 0.20 [0.1]\\ \hline
\hline
\end{tabular}
}
\caption{The priors and best-fits of the HOD parameters for the $\fastpm$ LRGs, ELGs, and QSOs, respectively. The HOD models vary for different tracers. We also show the corresponding HOD parameters of the $\abacus$ catalogs as the numbers in the square brackets.}  \label{tab:hodfit}
\end{table}
Based on the HOD fitting process described in section \ref{sec:method_hod}, we obtain the best-fit HOD parameters for the $\fastpm$ galaxy catalogs, including LRGs, ELGs and QSOs, as shown in table \ref{tab:hodfit}. We also show the priors of the fitting parameters for each model. Different tracers have different HOD models. Each blank space indicates that there is no such parameter in a specific HOD model. Note that there are significant differences on halo properties between the $\fastpm$ and $\abacus$ catalogs, such as halo mass function, we should not expect the fitted $\fastpm$ HOD parameters close to those of $\abacus$, even though we match the $\fastpm$ galaxy number density and two-point clustering at relatively large scales very well to those of $\abacus$. The $\fastpm$ HOD parameters do not have similar physical interpretation as the ones from normal $N$-body simulations.

\section{FastPM ELG and QSO clustering}
\begin{figure}
    \centering
\includegraphics[width=0.98\linewidth]{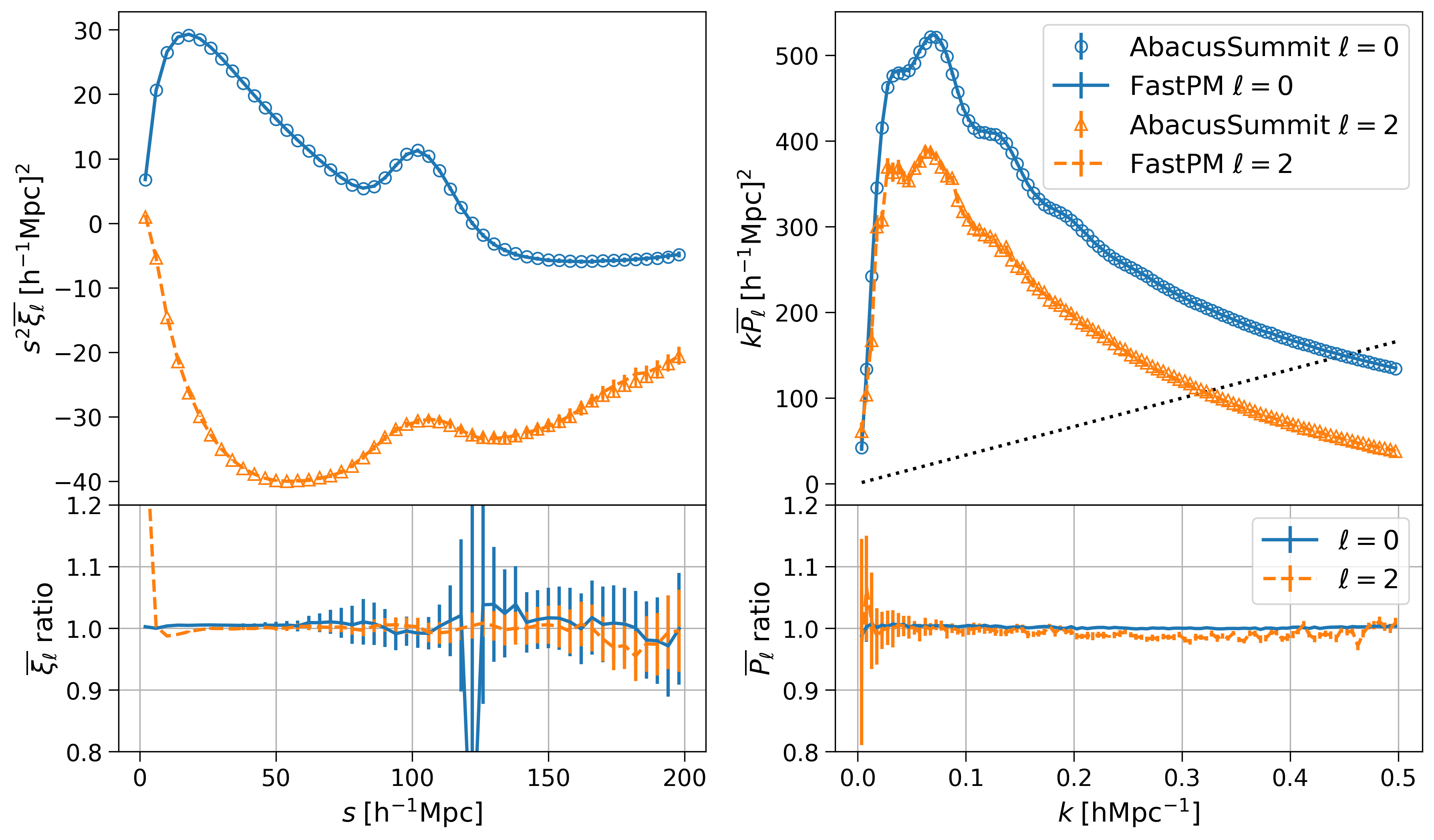}
    \caption{Same as figure \ref{fig:lrg_abacus_fastpm} but for the comparison of the ELG clustering from $\abacus$ and $\fastpm$.}
    \label{fig:elg_abacus_fastpm}
\end{figure}

\begin{figure}
    \centering
\includegraphics[width=0.98\linewidth]{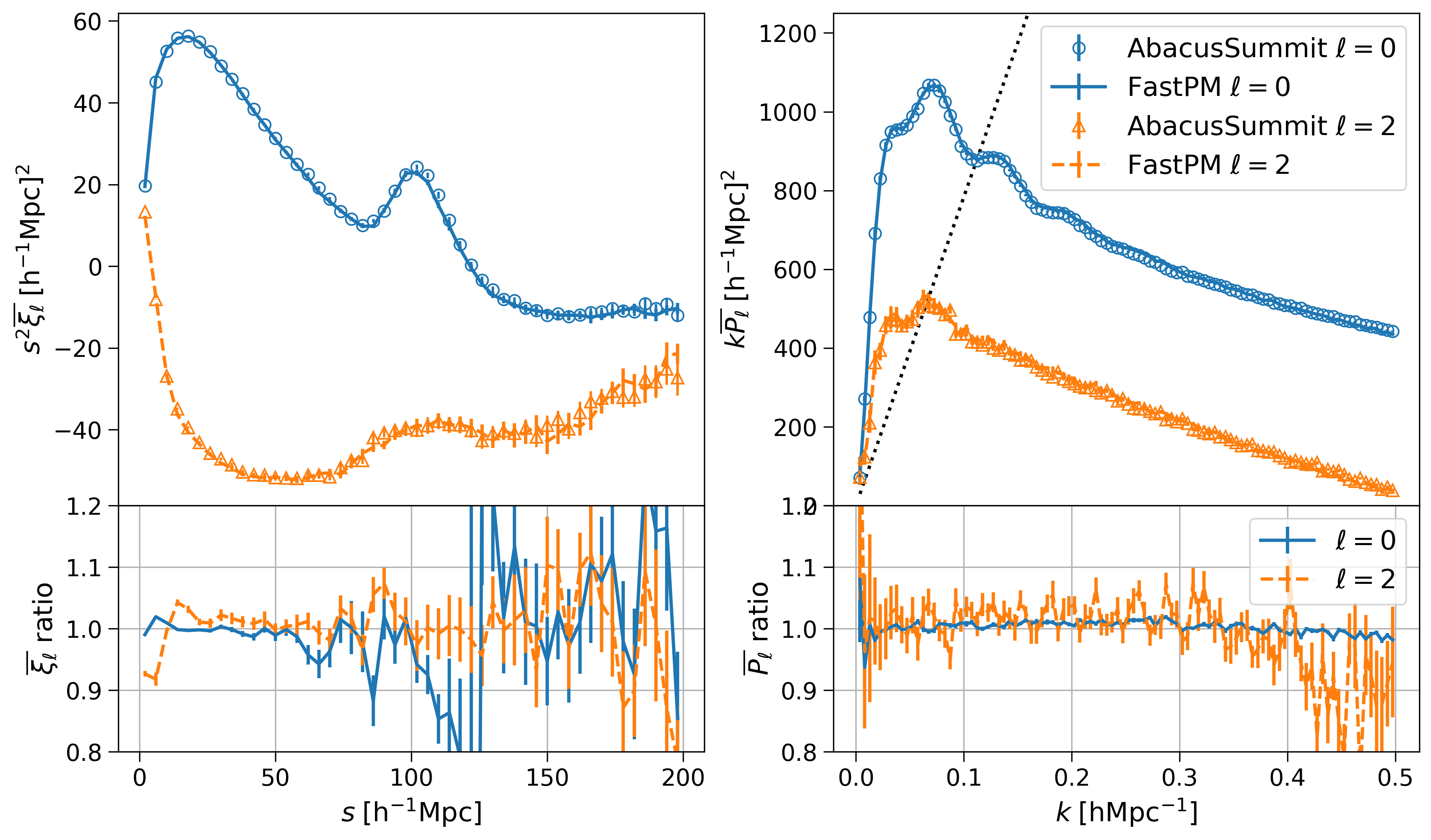}
    \caption{Same as figure \ref{fig:lrg_abacus_fastpm} but for the comparison of the QSO clustering from $\abacus$ and $\fastpm$.}
    \label{fig:qso_abacus_fastpm}
\end{figure}

With the best-fit HOD parameters (in table \ref{tab:hodfit}) for the $\fastpm$ ELGs and QSOs, we compare their two-point clustering with those of $\abacus$, as shown in figure \ref{fig:elg_abacus_fastpm}
and figure \ref{fig:qso_abacus_fastpm}. Same as figure \ref{fig:lrg_abacus_fastpm}, we display the mean over 25 catalogs. For QSOs, due to the low number density and large shot noise, the clustering signals are much noisier compared to those of LRGs and ELGs, shown as the larger fluctuation in the signal ratio. 

\begin{figure}
    \centering
\includegraphics[width=0.98\linewidth]{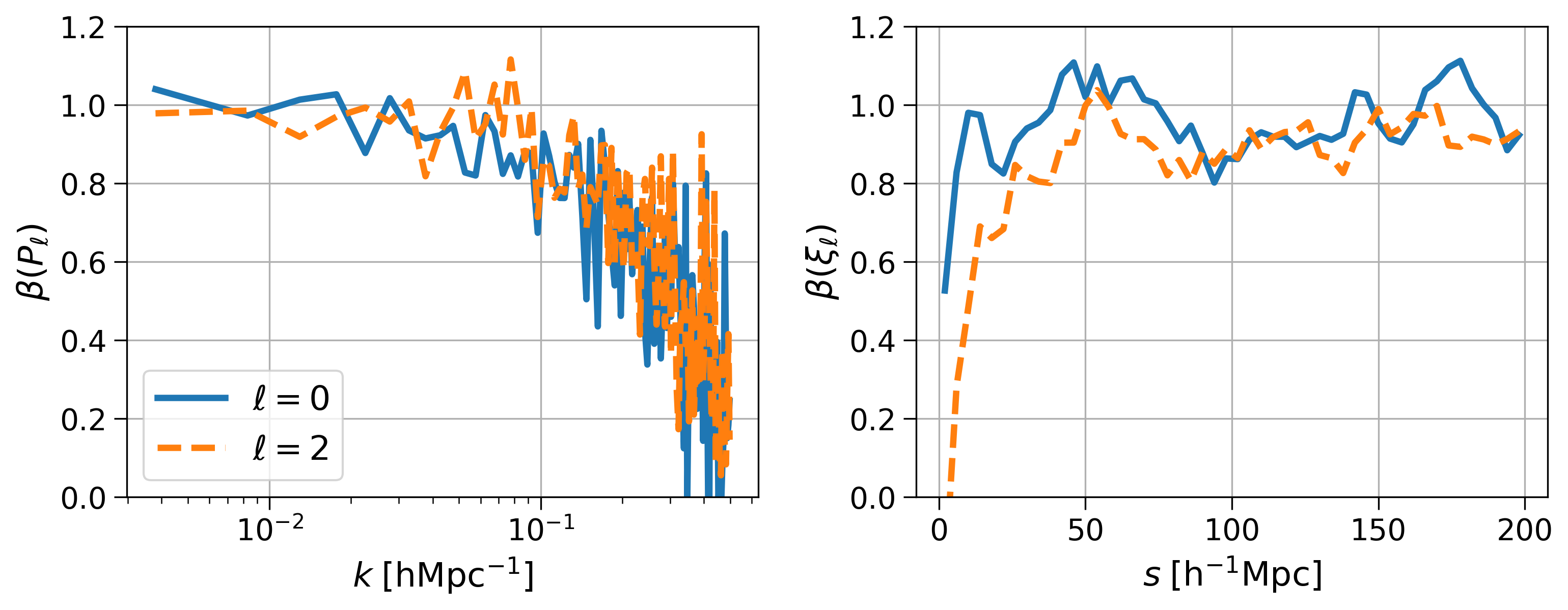}\\
\includegraphics[width=0.98\linewidth]{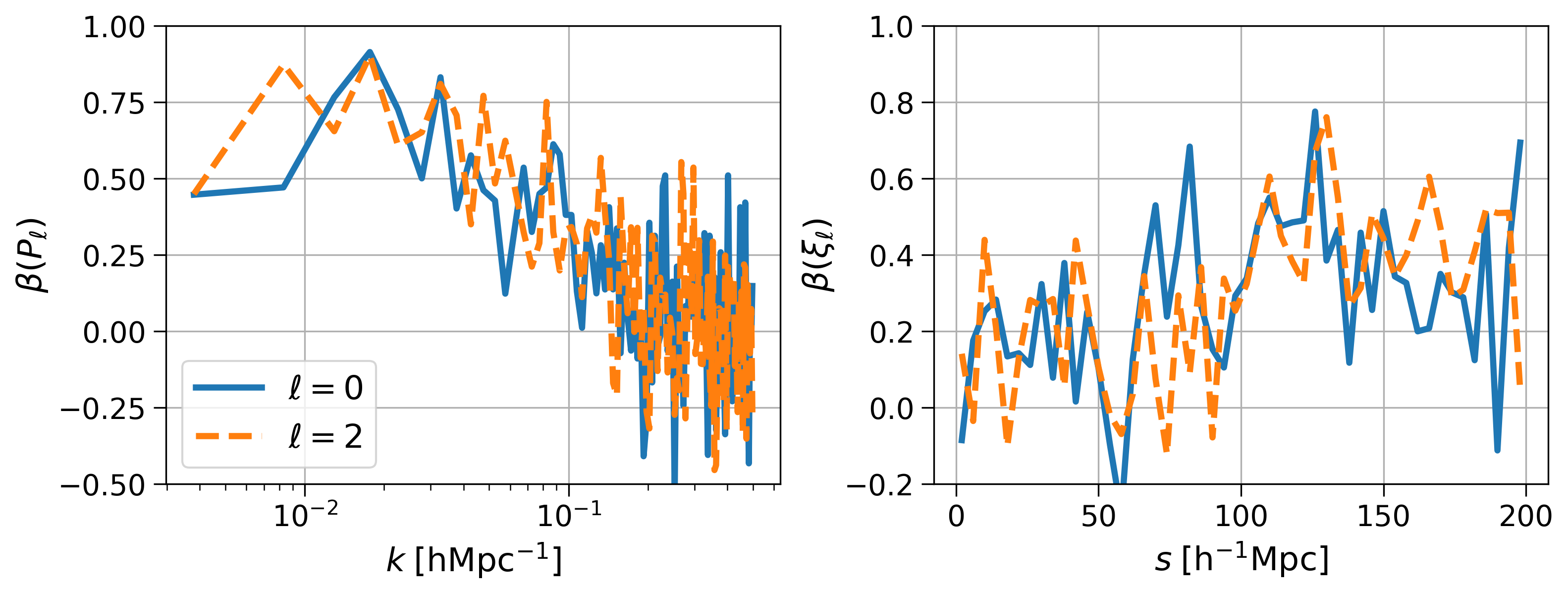}
    \caption{Similar to figure \ref{fig:lrg_beta}, we show the $\beta$ coefficients for the ELG and QSO clustering in the upper and lower panels, respectively.}
    \label{fig:elg_qso_beta}
\end{figure}

In figure \ref{fig:elg_qso_beta}, we also show the $\beta$ coefficients of the correlation function and power spectrum multipoles. The upper panels are for ELGs, and the lower panels are for QSOs. The $\beta$ value of QSOs is smaller than that of ELGs or LRGs, which we suspect is mainly due to the large shot noise. 
 
\section{BAO parameters $\alpha$ and $\epsilon$}
\begin{figure}
    \centering
\includegraphics[width=0.49\linewidth]{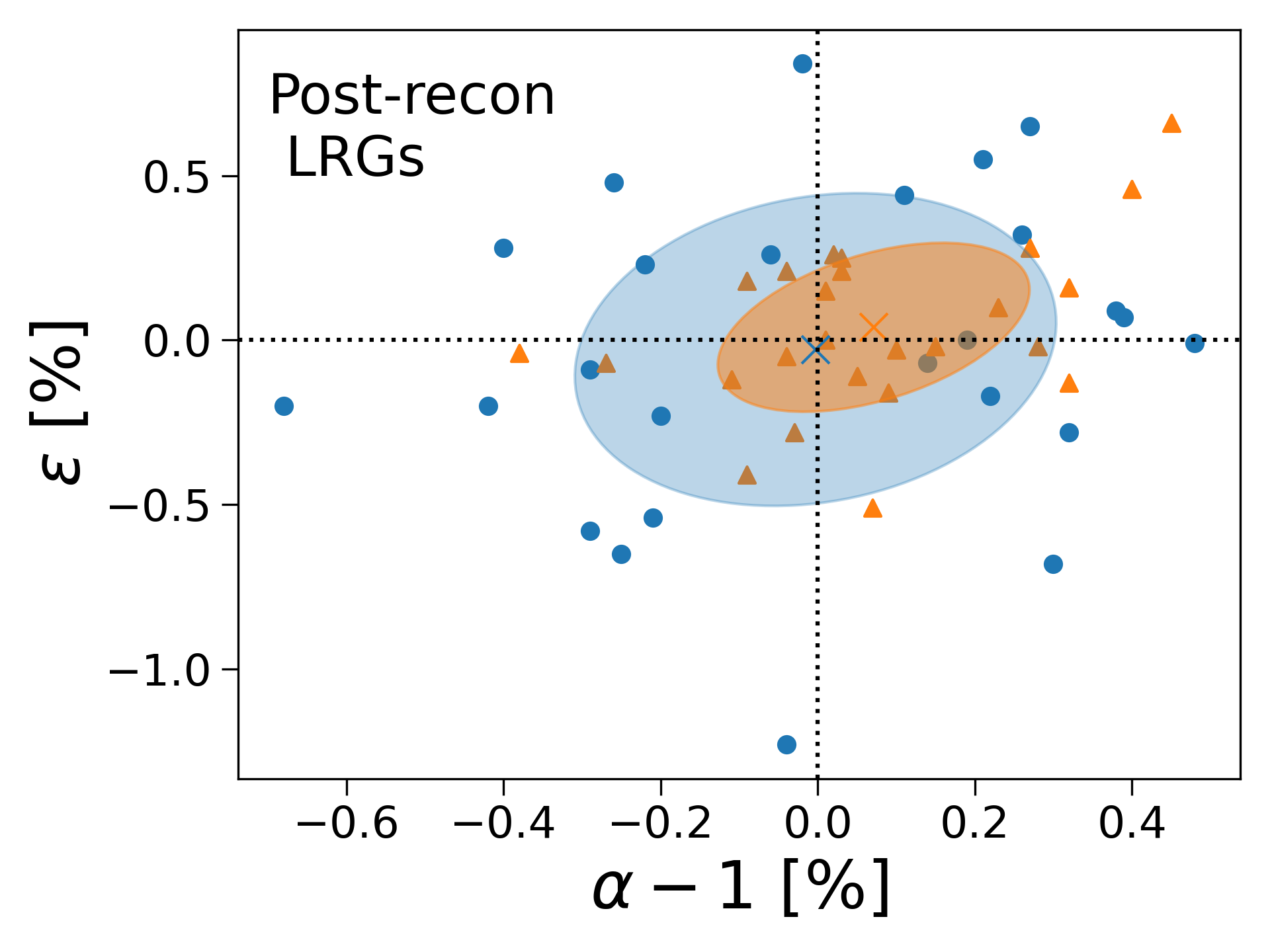}
\includegraphics[width=0.49\linewidth]{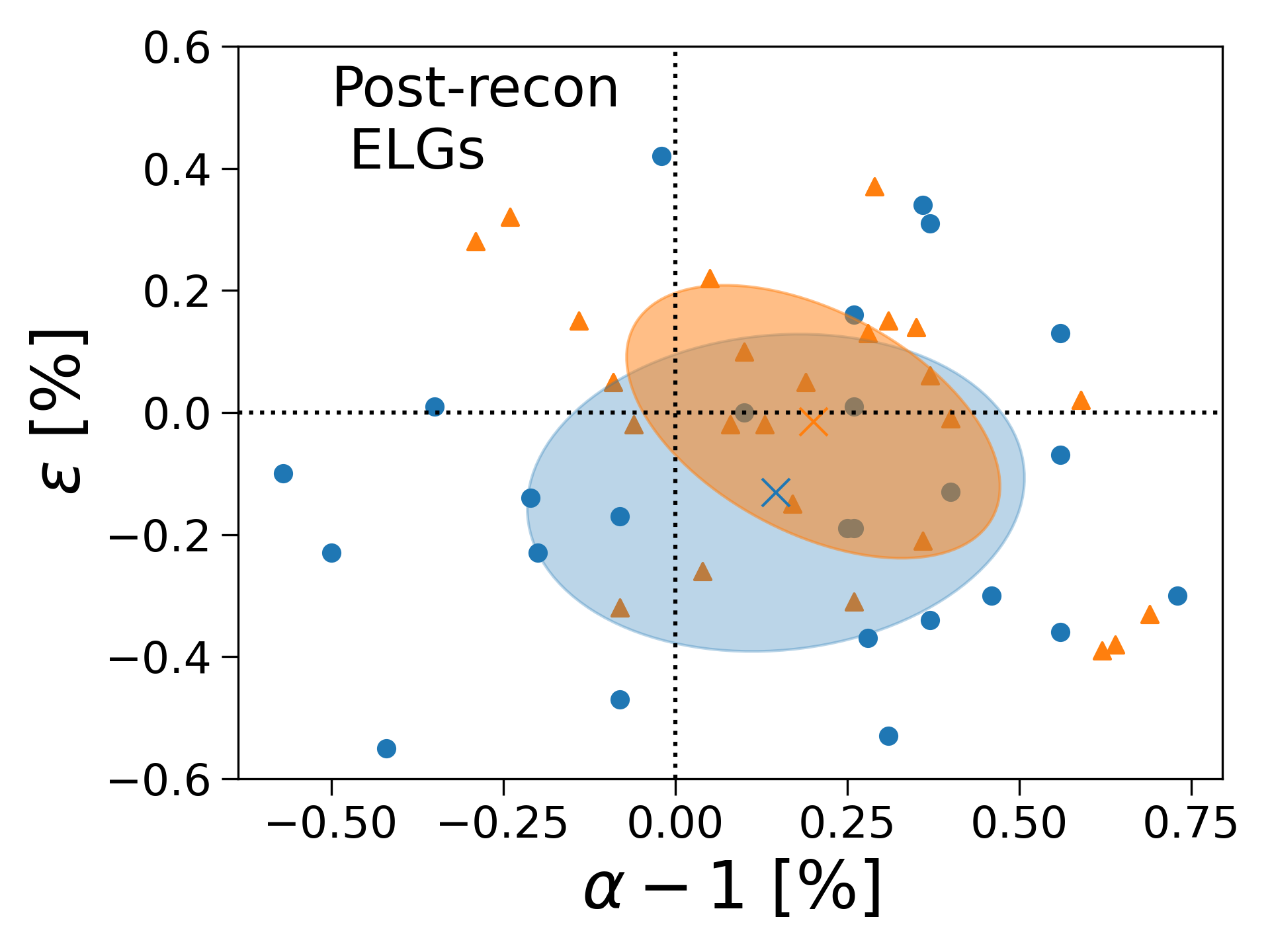}
    \caption{Similar to figure \ref{fig:alphas_lrg_elg} but for the fitting parameters $\alpha$ and $\epsilon$ (eq. \ref{eq:alpha} and \ref{eq:epsilon}). The left and right panels are for LRGs and ELGs, respectively. Both are after the BAO reconstruction. }
    \label{fig:alpha_epsilon}
\end{figure}

For complementary information, figure \ref{fig:alpha_epsilon} displays the isotropic and anisotropic BAO scale parameters $\alpha$ and $\epsilon$ for LRGs and ELGs over 25 realizations, respectively. We focus on the results after the BAO reconstruction. 

\section{ELG BAO fitting with the EZmock covariance}\label{sec:ezmock_cov}
\begin{figure}
    \centering
    \includegraphics[width=0.49\linewidth]{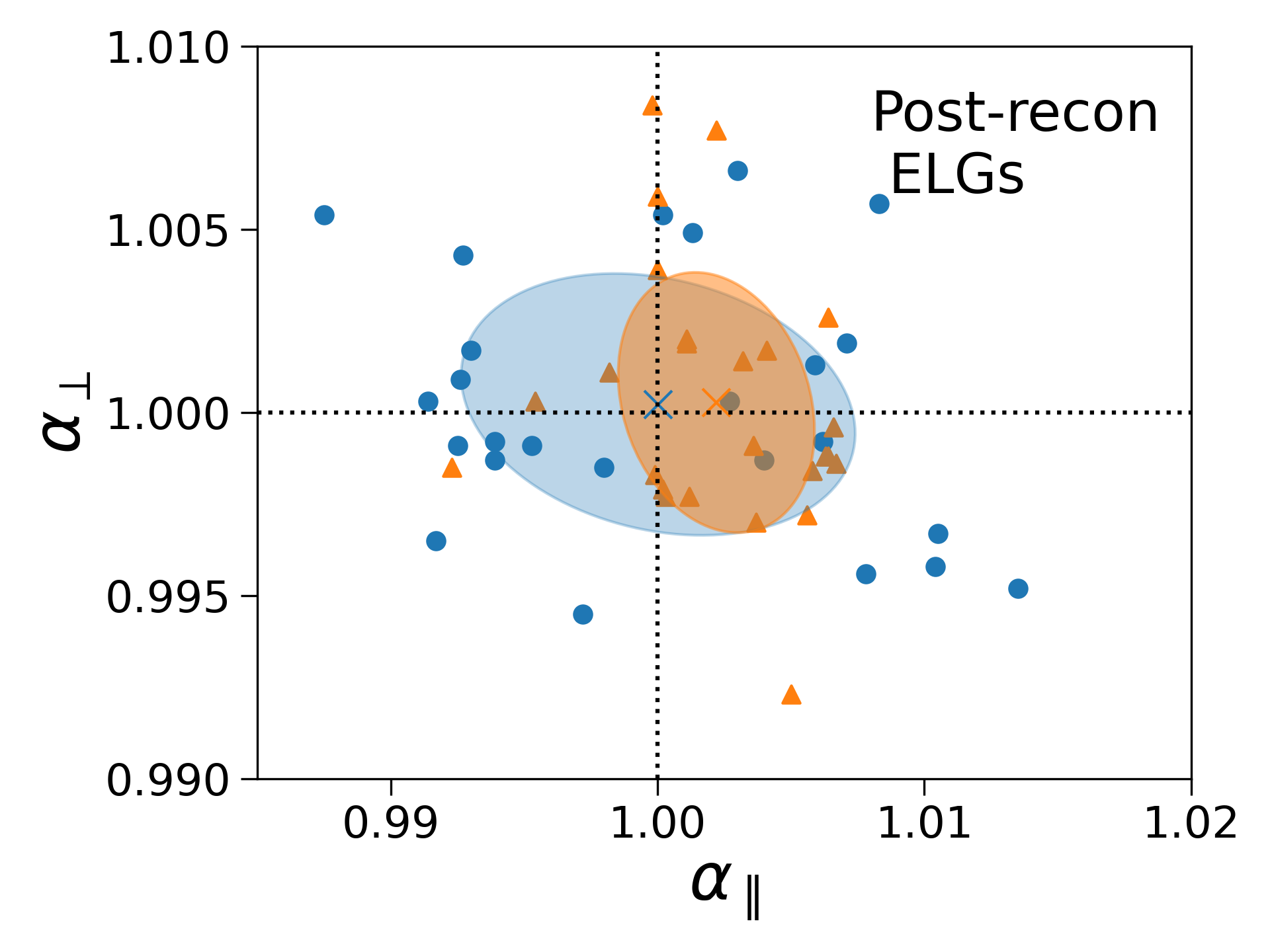}
    \includegraphics[width=0.49\linewidth]{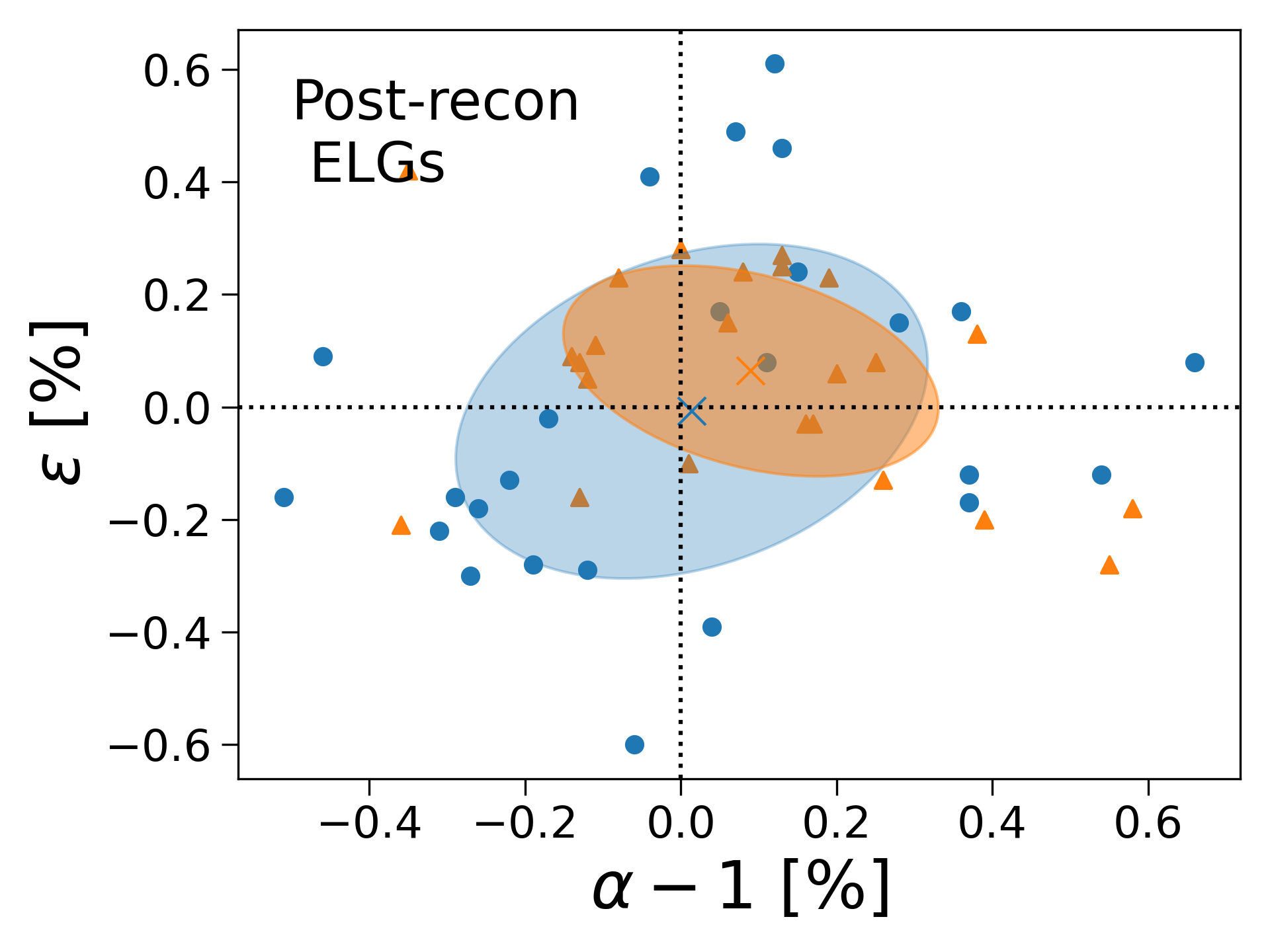}
    \caption{With the covariance matrix from $\textsc{EZmocks}$, we show the BAO fitting parameters for ELGs after the BAO reconstruction. The left (right) panel depicts the parameters $\alpara$ and $\alperp$ ($\alpha$ and $\epsilon$).}
    \label{fig:elg_bao_ezmockCov}
\end{figure}
From the lower right panel of figure \ref{fig:alphas_lrg_elg}, we notice that there is some bias on the BAO fitting parameter $\alpara$ for ELGs after the BAO reconstruction. We suspect that the systematics can be due to the $\fastpm$ covariance matrix, which is estimated from a limited number (313) of $\fastpm$ realizations. If we replace the covariance matrix by the one from 1000 $\textsc{EZmocks}$, which is adopted in recent studies \citep{Hadzhiyska2023b, KP4s11-Garcia-Quintero}, we can largely mitigate such systematics. We show the fitted BAO parameters in figure \ref{fig:elg_bao_ezmockCov}.

\section{Fixed-amplitude \textsc{FastPM} simulations}
\begin{figure}
    \centering
\includegraphics[width=0.98\linewidth]{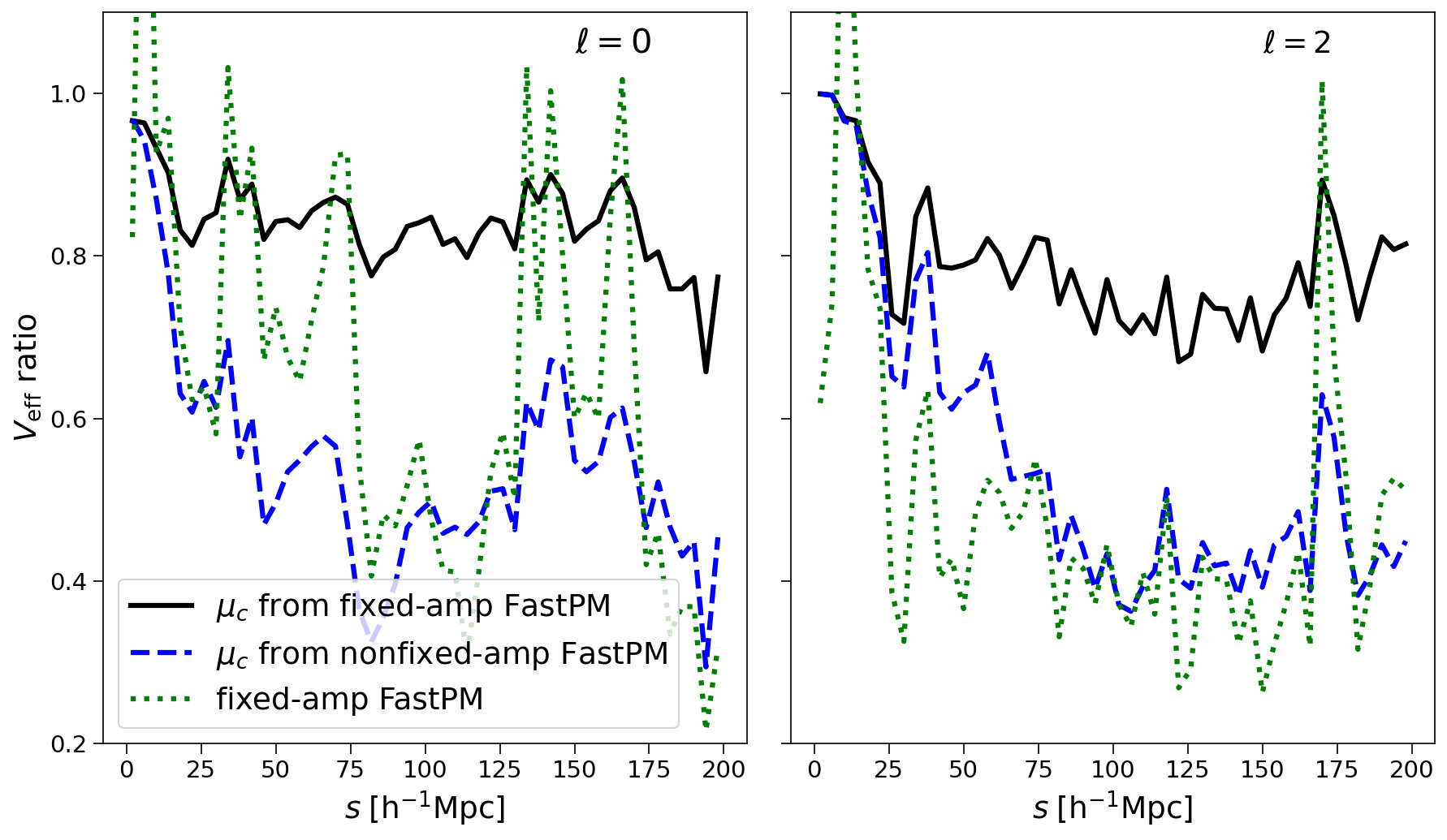}
    \caption{Influence of the $\mu_c$ error on the effective volume of 25 $\abacus$ LRG mocks from CARPool. We show the result for the correlation function monopole (left panel) and quadrupole (right panel). We consider the cases with $\mu_c$ estimated from a set of $\fastpm$ simulations with the fixed-amplitude and nonfixed-amplitude ICs, shown as the black solid and blue dashed lines, respectively. For each case, we calculate the ratio of the effective volume with respect to the optimal one with $\sigma^2_{\mu_c}=0$ (red lines in figure \ref{fig:err_veff_lrg}). In addition, we overplot the result if we use the fixed-amplitude $\fastpm$ simulations instead of CARPool to increase the effective volume, shown as the green dotted lines.}
    \label{fig:veff_ratio_lrg}
\end{figure}
It has been studied that using the fixed-amplitude ICs can effectively suppress the sample variance of dark matter and halo clustering, e.g. \citep{Chuang2019, Ding2022}. 
From the lower panels of figure \ref{fig:err_veff_lrg} and \ref{fig:err_veff_elg}, we have seen that the CARPool performance on the effective volume over multiple realizations depends on the sample variance of $\mu_c$, i.e. $\sigma^2_{\mu_c}$, which is estimated from a bunch of $\fastpm$ simulations in our study. We are interested in whether we can use the fixed-amplitude $\fastpm$ simulations to reduce the sample variance of galaxy clustering, hence, we can improve the CARPool performance.

In our previous work \citep{Ding2022}, we have generated more than $200$ fixed-amplitude $\fastpm$ halo catalogs. As an instance, we populate LRGs in the fixed-amplitude simulations adopting the same HOD parameters as the non-fixed-amplitude $\fastpm$ at $z=0.8$. We calculate the correlation function multipoles and the covariance matrix over 200 fixed-amplitude realizations. In figure \ref{fig:veff_ratio_lrg}, we compare the effective volumes of 25 $\abacus$ realizations from the cases with $\mu_c$ estimated from the default (non-fixed-amplitude) and fixed-amplitude $\fastpm$ simulations, respectively. Note that for the non-fixed-amplitude case, we use 313 realizations. 
To show the relative difference, we have rescaled the volumes by the reference, which is the optimal CARPool result with $\sigma^2_{\mu_c}=0$, i.e. assuming no sample variance on $\mu_c$. With smaller $\sigma_{\mu_c}$ from the fixed-amplitude $\fastpm$, $V_{\text{eff}}$ increases significantly and reaches $\sim 80$ per cent of the optimal one. We also compare the volume of 25 fixed-amplitude $\fastpm$ LRGs with respect to the optimal CARPool result, shown as the green dotted lines. Interestingly, it is comparable with the CARPool result with $\mu_c$ from the non-fixed-amplitude $\fastpm$. This test also demonstrates the power suppressing the sample variance via combining the fixed-amplitude and CARPool methods.